\documentclass{ametsocV6.1}

\usepackage{gensymb}
\usepackage{xcolor}
\usepackage{amsmath}

\title{Leveraging data-driven weather models for improving numerical weather prediction skill through large-scale spectral nudging}

\authors{Syed Zahid Husain\aff{a},\correspondingauthor{Syed Zahid Husain, syed.husain@ec.gc.ca}
Leo Separovic\aff{a},
Jean-Fran\c cois Caron\aff{b},
Rabah Aider\aff{a},
Mark Buehner\aff{b},
St\'ephane Chamberland\aff{a},
Ervig Lapalme\aff{b},
Ron McTaggart-Cowan\aff{a},
Christopher Subich\aff{a},
Paul A. Vaillancourt\aff{a},
Jing Yang\aff{a},
and Ayrton Zadra\aff{a}
}

\affiliation{\aff{a}{Atmospheric Numerical Prediction Research Section, Environment and Climate Change Canada, Dorval, Quebec, Canada H9P 1J3}\\
\aff{b}{Data Assimilation and Satellite Meteorology Research Section, Environment and Climate Change Canada, Dorval, Quebec, Canada H9P 1J3}\\
}

\abstract{Operational meteorological forecasting has long relied on physics-based numerical weather prediction (NWP) models. Recently, this landscape has faced disruption by the advent of data-driven artificial intelligence (AI)-based weather models, which offer tremendous computational performance and competitive forecasting accuracy. However, data-driven models for medium-range forecasting generally suffer from major limitations, including low effective resolution and a narrow range of predicted variables. This study illustrates the relative strengths and weaknesses of these competing paradigms using the physics-based GEM (Global Environmental Multiscale) and the AI-based GraphCast models. Analyses of their respective global predictions in physical and spectral space reveal that GraphCast-predicted large scales outperform GEM, particularly for longer lead times, even though fine scales predicted by GraphCast suffer from excessive smoothing. Building on this insight, a hybrid NWP-AI system is proposed, wherein temperature and horizontal wind components predicted by GEM are spectrally nudged toward GraphCast predictions at large scales, while GEM itself freely generates the fine-scale details critical for local predictability and weather extremes. This hybrid approach is capable of leveraging the strengths of GraphCast to enhance the prediction skill of the GEM model while generating a full suite of physically consistent forecast fields with a full power spectrum. Additionally, trajectories of tropical cyclones are predicted with enhanced accuracy without significant changes in intensity. Work is in progress for operationalization of this hybrid system at the Canadian Meteorological Centre.}

\begin{document}

\maketitle

%
\section{Introduction}
State-of-the-art physics-based NWP models include some form of a dynamical core that solves the atmospheric governing equations, and are coupled to a suite of parameterization schemes to represent diabatic, frictional, and subgrid-scale processes that are not explicitly accounted for by the dynamical equations. Although statistical alternatives have been explored in the past --- particularly for downscaling purposes \citep{ybg06, btc08, clf14} --- physics-based models have long been the foundational approach for operational meteorological forecasting. The recent emergence of data-driven models inspired by artificial intelligence (AI) has, however, started to seriously challenge this well established paradigm.

\cite{kei22} presented an important breakthrough demonstrating considerable potential for weather forecasting with data-driven models. This was quickly followed by Pangu-Weather from Huawei \citep{bxz22,bxz23}, GraphCast from Google DeepMind \citep{lsw22,lsw23}, and several other models (e.g., \citealt{psh22, chg23, czz23}). In general, these models rely on some form of deep neural network architecture. Like any other application of AI, data-driven weather models require a substantial volume of high quality training data. As a result, all currently available AI-based deterministic global weather simulators are trained on the ERA5 reanalyses \citep{hbb20} from the European Centre for Medium-Range Weather Forecasts (ECMWF), which is undoubtedly the most comprehensive resource available. The neural network weights within these data-driven models are specifically trained to make inferences (forecasts) that closely emulate ERA5.

Although training of AI-based weather emulators is computationally expensive, their exceptional computational performance during inference accelerates production times by orders of magnitude while using a fraction of the computational resources usually devoted to the physics-based models. In addition to their efficiency, standard headline scores suggest that data-driven systems generate predictions that are more accurate even than those of the Integrated Forecasting System (IFS; \citealt{lrs23}) from ECMWF, the model used to generate the ERA reanalyses on which data-driven models were trained.

Despite these advantages, AI-based weather emulators also have limitations. A major weakness of most of the currently available data-driven models is considerable fine-scale smoothing \citep{bxz22,lsw22}. Furthermore, these models are unable to accurately represent the fundamental dynamical balances in the atmosphere, leading to smoothing that is inconsistent across physically-related variables \citep{bon24}. It is claimed that smoothing can largely be addressed by  employing a diffusion architecture \citep{psa24}; however, such approaches currently increase computational cost considerably, thereby diminishing one of the principal advantages of the AI paradigm.

Increasing the nominal horizontal resolution of AI inferences will also be challenging, as it would require higher resolution training data. Since ERA5 reanalyses are only available on a 0.25$\degree$ grid, current global AI models cannot be trained for kilometer-scale forecasting. Current data-driven systems are also only capable of predicting a limited number of variables, a subset of those available in the training dataset. Training to predict non-analyzed variables (e.g., precipitation in ERA5) generally leads to suboptimal inference \citep{lsw22,lsw23}. Predicting the full suite of physics-related prognostic variables required of operational NWP, such as categorization of cloud and precipitation types, would also substantially increase the computational cost. Time resolution of inferences is another constraint for most of the current AI models, whose 6-hour prediction intervals are far coarser than physics-based equivalents. Pangu-Weather \citep{bxz22} provides separate models for 1-, 3-, 6-, and 24-hour inferences, but the models with shorter forecasting steps suffer from larger error accumulation. These limitations make it impossible for current AI-based models to completely replace operational physics-based NWP systems.

The ultimate goal of this study is to develop a hybrid NWP-AI system for real-time global forecasting applications that combines the strengths of each paradigm while overcoming their individual limitations. Although the techniques developed here are generally applicable, the implementation presented in this paper is based on Environment and Climate Change Canada's (ECCC's) physics-based Global Environmental Multiscale (GEM) model \citep{gdm14} and the AI-based GraphCast model from Google DeepMind.

The hybridization involves large-scale spectral nudging of GEM predictions toward GraphCast inferences. Although the concept of spectral nudging was first proposed to control spatial computational modes in a limited-area model (\citealt{wph96}), it is primarily used to improve dynamical downscaling over high-resolution sub-domains for regional climate modelling \citep{lle09,hsy14}. In general, it can be interpreted as an indirect suboptimal data assimilation method \citep{slf00}. Spectral nudging has also been explored for global atmospheric hindcasting \citep{sff17}.

Spectral nudging has so far been unfeasible for real-time global forecasting applications due to the lack of a timely and accurate reference data source. However, the emergence of AI models, with their rapid inference speed and improved large-scale skill, as shown in this study, has unlocked the potential for applying spectral nudging in real-time operational global weather forecasting.

The development of the hybrid NWP-AI system presented here, begins with background information on the GEM and GraphCast models in section \ref{s_models}. The relative performance of these models is assessed in section \ref{s_rp}. Section \ref{s_spn} introduces spectral nudging in GEM and documents an optimal nudging configuration. Detailed evaluations of the hybrid system are presented in section \ref{s_eva}, followed by a summary of the study's main conclusions and recommendations for future work in section \ref{s_summ}.

\section{Model descriptions}
\label{s_models}
\subsection{The GEM Model}
The dynamical core of the GEM model solves the elastic Euler system of equations using an iteratively implicit semi-Lagrangian approach \citep{gdm14,hgi17}. The equations are first transformed from regular height coordinate to some form of a terrain-following coordinate in the vertical, denoted by $\zeta$ in the model, which for ECCC's current operational NWP systems, is based on log-hydrostatic-pressure \citep{hgs21}. The global domain is represented as a pair of overlapping limited-area grids in a Yin-Yang configuration \citep{qle11}. Solutions from the GEM dynamical core are combined with tendencies from parameterized physics schemes \citep{mcc19} that represent the diabatic, frictional, and subgrid-scale processes. The resulting tendencies augment the dry-dynamical solution via sequential/split coupling \citep{gwr18,hgq19} to produce the complete solution for a model time step.

The GEM configuration used here is that of the current operational GDPS (Global Deterministic Prediction System; \citealt{mcc19,gdps9}), with a horizontal grid spacing of approximately 15 km and 450 s time step. The GDPS has 84 prognostic vertical levels with the top approximately at 0.1 hPa.

Although operational GDPS is coupled with the NEMO ocean model \citep{sbr18}, all the experiments for this study were conducted with atmosphere-only configurations. This simplification reduces both complexity and computational cost, and in the past, has been found to provide reliable guidance for the behavior of the full operational system.

\subsection{The GraphCast Model}
The AI-based GraphCast model from Google DeepMind has been trained to emulate ECMWF's ERA5 reanalyses available with a horizontal grid spacing of 0.25$\degree$ \citep{lsw22,lsw23}. It is trained to minimize the mean squared error (weighted by vertical levels) for up to 12 recursive forecast steps (from 6 hr to 3 days), and is capable of producing forecasts with reasonable accuracy up to day 10. The atmospheric state predicted by GraphCast is represented by 6 variables defined on multiple pressure levels (temperature, $u-v$ components of wind, geopotential, specific humidity, and vertical wind speed) and 5 surface variables (2-m temperature, 10-m wind components, mean sea-level pressure, and total precipitation).

At its core, GraphCast employs graph neural networks (GNNs) with an ``encoder-processor-decoder'' configuration \citep{lsw22,lsw23}. The input and output states of GraphCast are represented over a 0.25$\degree$ latitude-longitude global grid. A major advantage of GNNs is the possibility of having arbitrary range of spatial interactions. GraphCast takes advantage of this GNN feature by introducing a multi-mesh architecture based on iteratively-refined icosahedral grids (from level-0 to level-6 refinements) within the processor component. The encoder of GraphCast maps data from the input 0.25$\degree$ latitude-longitude grid to the internal multi-mesh of the processor, whereas the decoder brings back information from the processor multi-mesh to the output grid.

This study employs the 13 pressure-level version of GraphCast with pre-trained weights (learned features of the GNNs) that are available from Google DeepMind. Although a 37-level version is available, only the 13-level variant has been subjected to additional fine-tuning with ECMWF's operational analyses (2016–2021), making it more skillful than the 37-level version. In this study, GraphCast is initialized with operational GDPS analyses \citep{cbu22} without any fine-tuning applied to its weights.

\section{Relative performances of the GDPS and GraphCast forecasts}
\label{s_rp}
Ten-day forecasts from GDPS and GraphCast are initialized every 36 h, leading to 60 cases for boreal winter 2022 (from 01 January 2022 at 1200 UTC to 31 March 2022 at 0000 UTC) and 62 cases for boreal summer 2022 (from 01 June 2022 at 0000 UTC to 31 August 2022 at 1200 UTC). The sequence of forecast integrations initialized from these dates using an atmosphere-only operational GDPS configuration is denoted GDPS-CTL.

Predictions from the two systems (GDPS-CTL and GraphCast) are first assessed for their ability to depict the full range of atmospheric circulations using variance spectra (section \ref{s_rp}a). This analysis is aimed at ascertaining effective resolutions of the two models, which will help to fairly interpret their relative predictive skills when comparing against radiosonde observations later in section \ref{s_rp}b.

\subsection{Verification with global power spectra}
\begin{figure}
\begin{center}
\noindent\includegraphics[width=38pc,angle=0]{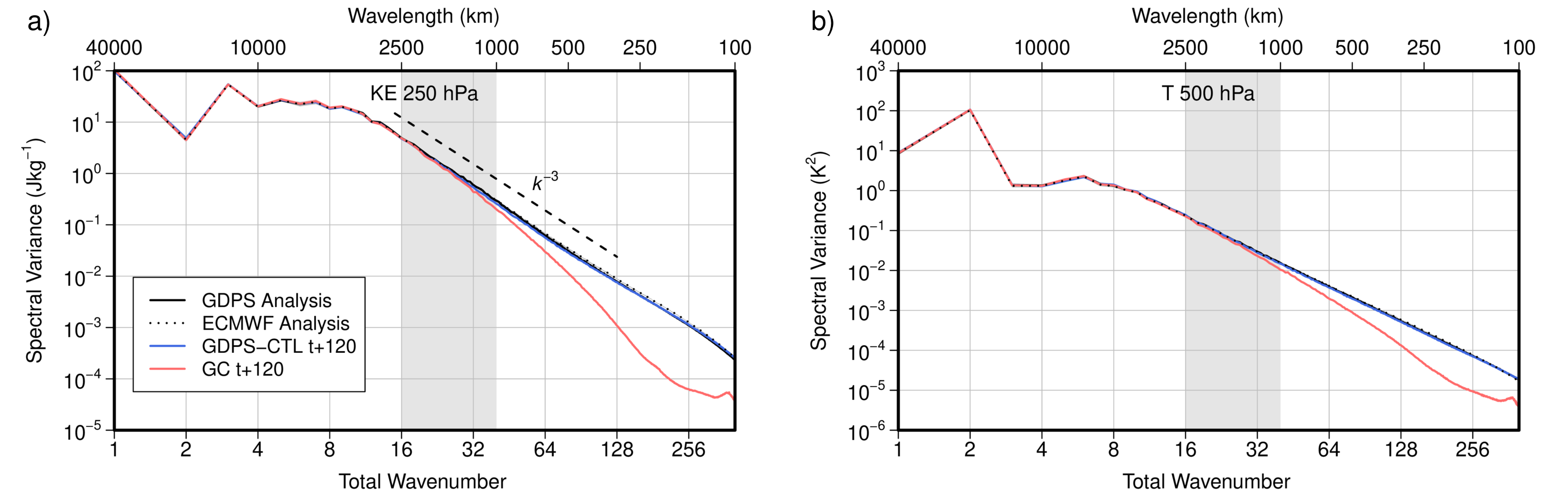}\\
\caption{Global (a) kinetic energy power spectrum of 250-hPa horizontal wind and (b) variance spectrum of 500-hPa temperature for the GDPS analyses (black) and ECMWF analyses (dotted black) as well as 120-hr lead time of GDPS-CTL (blue) and GraphCast (red), averaged over 60 cases of boreal winter 2022. The dashed line indicates the -3 slope. The 5\%--95\% confidence interval is presented with light shadings (not discernible for the given logarithmic scale). The small- and large-scale cut-offs for the global filter (appendix B) for fair comparison at low resolution (section \ref{s_rp}b2) are approximately 1000 km and 2500 km, respectively, as indicated with the shaded area.}
\label{f_tot_spc_1}
\end{center}
\end{figure}

Although GraphCast employs a 0.25$\degree$ latitude-longitude resolution grid, its predictions are subject to a smoothing effect that increases with lead time. This is believed to be associated with learning to minimize the mean square error (MSE) (e.g., \cite{kei22}, \cite{lsw23}). The smoothing effect is evidenced by departures of 120-hr GraphCast forecasts’ global 250-hPa kinetic energy and 500-hPa temperature variance spectra from those of GDPS-CTL forecasts, as well as GDPS and ECMWF analyses (Fig. \ref{f_tot_spc_1}). For the fine scales, GraphCast’s spectra are steeper, leading to a notable variance deficiency at wavenumbers as low as 30, corresponding to lower synoptic and sub-synoptic scales ($<$1500 km). Unlike GraphCast, both GDPS-CTL forecasts and the two analyses closely follow the expected $k^{-3}$ spectral slope for 3D quasi-geostrophic flow \citep{cha71,spk14}. The fact that the spectral variance of ECMWF analyses -- on which GraphCast has been fine-tuned -- closely matches that of both GDPS-CTL forecasts and GDPS analyses demonstrates that GraphCast's fine-scale smoothing does not originate from any lack of fine-scale variance in the training data. Overall, the effective resolution, defined as the highest wave number at which the atmospheric variability is still adequately represented, is considerably lower for GraphCast than its nominal resolution, i.e., grid spacing. Conversely, the effective resolution of the classical NWP models is typically about 6$\sim$8 times their nominal resolution \citep{ska04}. With the current operational configuration, the nominal resolution for GDPS-CTL is 0.1375$\degree$, leading to an approximate effective resolution of 90$\sim$120 km \citep{mca24}.

It is important to note that the stationary effects of orography, land-sea contrast and latitudinal climatological variations tend to be much less affected by smoothing in GraphCast than the less-predictable transient anomalies. Let $x$ and $y$ denote the forecast and analysis, respectively. The transient anomalies -- which indicate deviations of individual forecasts (or analyses) from the seasonal mean state -- are then defined as $x'=x-\overline{x}$ and $y'=y-\overline{y}$, where the overbar denotes an average over all cases within the season of interest. It is then useful to introduce the transient spectral amplitude ratio, defined as follows:

\begin{equation}
\label{e_var}
 \gamma(n) = \left(\frac{\overline{\sigma_{x',x'}(n)}}{\overline{\sigma_{y',y'}(n)}}\right)^{1/2},
\end{equation}

\noindent where $\sigma_{x',x'}$ and $\sigma_{y',y'}$ are real-valued non-negative functions of the total spherical wavenumber $n$, representing global spectral variances of the transient-eddy components associated with the forecast and analysis, respectively. The optimal value of $\gamma$ is 1 for the range of scales that are adequately represented in the reference analysis. In general, for any given length scale (or wavenumber), a value of $\gamma$ larger than 1 would imply that the forecast has more variability than the analysis, whereas a value smaller than 1 would indicate variance deficiency in the forecast. By replacing the seasonal means with climatology and integrating the individual terms on the right-hand side of Eq. \ref{e_var} across the entire spectrum of wavenumbers, one may obtain the ratio of forecast to analysis activity. Further information on forecast activity and its spectral decomposition is provided in appendix A.

Figures \ref{f_var_coh_1}a,b show the evolution of $\gamma$ with lead time for GraphCast and GDPS-CTL forecasts against the GDPS analysis. A model's effective resolution can be obtained by identifying the smallest wavenumber at which it suffers from a considerable drop in $\gamma$. A value of $\gamma<0.9$, implies an amplitude damping of 10\%, and is assumed to be the threshold for determining a model's effective resolution in this study. While $\gamma$ remains close to 1 for all lead times for GDPS-CTL, indicating that its effective resolution does not change during the integration, for GraphCast it rapidly decreases, leading to an effective resolution as low as 1000 km for 24-hr forecasts. The smoothing effect further reduces the effective resolution to 2500$\sim$2700 km before it saturates around forecast days 3$\sim$5. It is worth noting here that GraphCast still resolves scales smaller than these limits, but only partly ($\gamma<0.9$). The impact of this variance-deficiency on optimal usage of GraphCast guidance in a hybrid system will be discussed in more detail in section \ref{s_spn}.

\begin{figure}
\begin{center}
\noindent\includegraphics[width=38pc,angle=0]{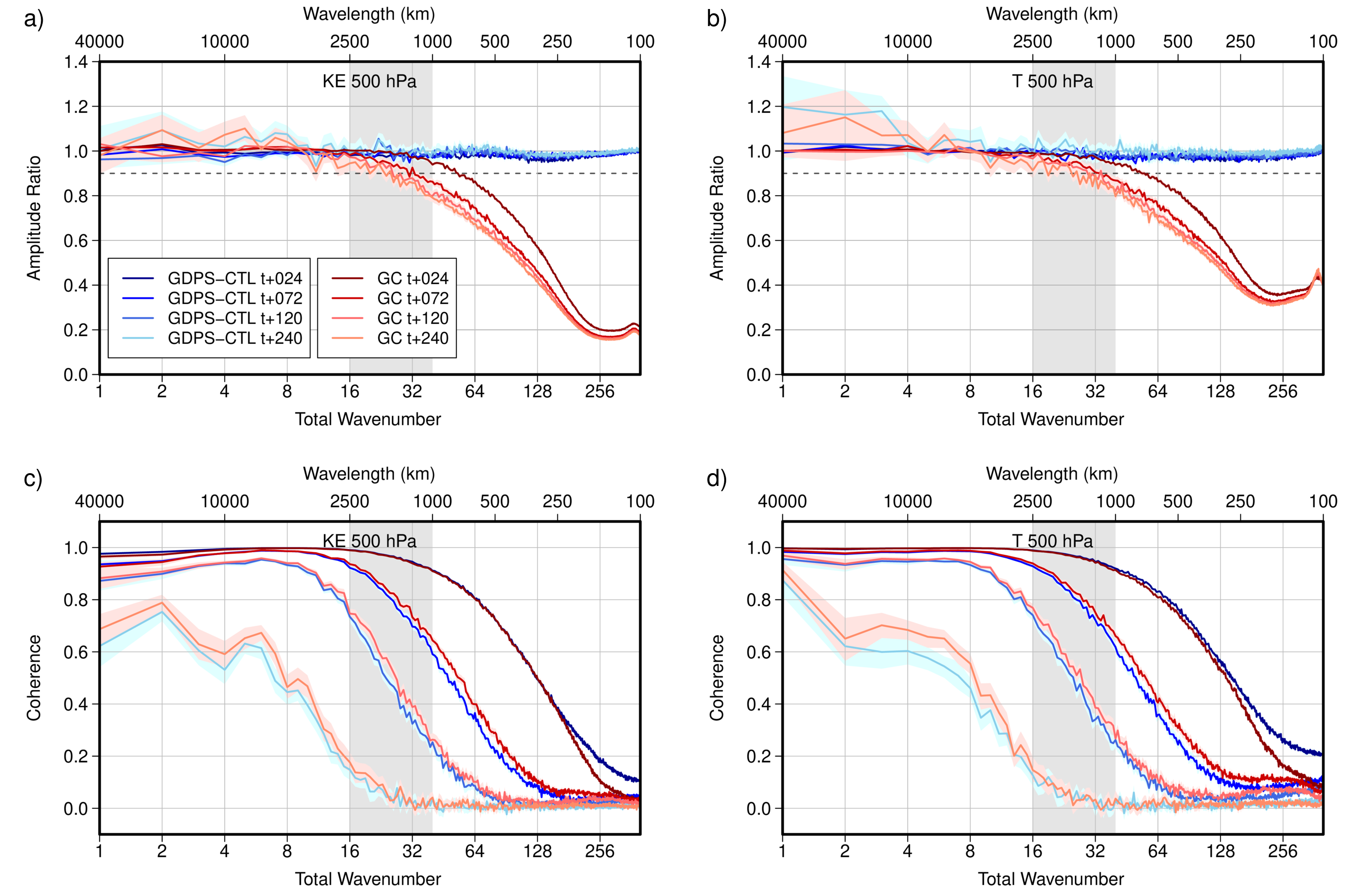}\\
\caption{Spectral (a, b) amplitude ratio $\gamma$, and (c, d) coherence $\rho$ for the global (a, c) kinetic energy spectrum of 500-hPa horizontal wind transient-eddy deviations, and (b, d) variance spectrum of 500-hPa temperature transient-eddy deviations, for GDPS-CTL (blue) and for GraphCast (red), after 24, 72, 120, and 240 hr of integration (darker to lighter shades of color). Spectra are averaged over 60 cases of boreal winter 2022. Light shadings indicate the 5\%--95\% confidence interval. The dashed horizontal lines in Figs. \ref{f_var_coh_1}a,b indicate $\gamma$=0.9. The small- and large-scale cut-offs for the global filter (appendix B) for fair comparison at low resolution (section \ref{s_rp}b2) are approximately 1000 km and 2500 km, respectively, as indicated with the shaded area.}
\label{f_var_coh_1}
\end{center}
\end{figure}

Figures \ref{f_var_coh_1}c,d display the spectral coherence defined as follows:
\begin{equation}
\label{e_coh}
 \rho(n) = \frac{\overline{\sigma_{x',y'}(n)}}{\left(\overline{\sigma_{x',x'}(n)}\;\overline{\sigma_{y',y'}(n)}\right)^{1/2}}\;,
\end{equation}

\noindent where $\sigma_{x',y'}$ is a real-valued function of the total spherical wavenumber $n$, representing the transient-eddy covariance spectrum between the GDPS-CTL (or GraphCast forecasts) and the GDPS analyses. Spectral coherence can be interpreted as a scale-dependent spatial correlation coefficient with an optimal value of 1, which would imply a perfect spatial correlation between forecast and analysis transient-eddy anomalies. Conversely, a value of 0 would imply a total de-correlation. By replacing the seasonal means with climatology and integrating the individual terms on the right-hand side of Eq. \ref{e_coh} across the entire spectrum of wavenumbers, one may obtain a quantity which is analogous to the anomaly correlation. Further information in this regard is presented in appendix A.

Since the same GEM model is used to produce both the GDPS forecasts and backgrounds for the analysis steps, they may have common systematic errors, especially at shorter lead times. Even though this penalizes the ERA5-trained GraphCast, Figs. \ref{f_var_coh_1}c-d indicate that GraphCast long-lead inferences are more skillful compared to the equivalent GDPS predictions over a broad range of synoptic and planetary scales.

\subsection{Verification against radiosondes}
At ECCC, model developers often rely on forecast verification against ECMWF analysis \citep{lrs23} to avoid the problem associated with own-analysis verification (see, e.g., \cite{cws08}, \cite{pet21}). Conversely, GraphCast's training likely introduces some dependency on the ECMWF analysis system. To ensure that the dataset used to verify the forecasts is as independent as possible from both models as well as the GDPS data assimilation system, it was opted to mainly rely on observations from the global radiosondes network for zonal wind ($U$), wind modulus ($|\mathbf{V_H}|$), geopotential height ($Z$), temperature ($T$), and dewpoint depression ($T–T_d$; where $T_d$ denotes dewpoint temperature) at mandatory pressure levels between 1000 and 100 hPa.

Two flavors of verification against radiosonde observations are performed based on the results shown in section \ref{s_rp}a: (1) verification at full resolution, with the GDPS-CTL and GraphCast forecasts at their native resolutions, and (2) verification at a low resolution –- corresponding to the effective resolution of GraphCast (around day 3) -- wherein both the GDPS-CTL and GraphCast forecasts are filtered to remove the variability at scales not adequately resolved by GraphCast. The latter is required because GraphCast's smoothing artificially reduces the ``double penalty'' for misplaced patterns in the nominal resolution evaluation \citep{mow02}. Low resolution fields are computed using a spectral filter based on spherical harmonics. The response function of the filter reflects the effective resolution of GraphCast at forecast days 3 to 5 (Figs. 2a and b; details in appendix B).

\begin{figure}
\begin{center}
\noindent\includegraphics[width=33pc,angle=0]{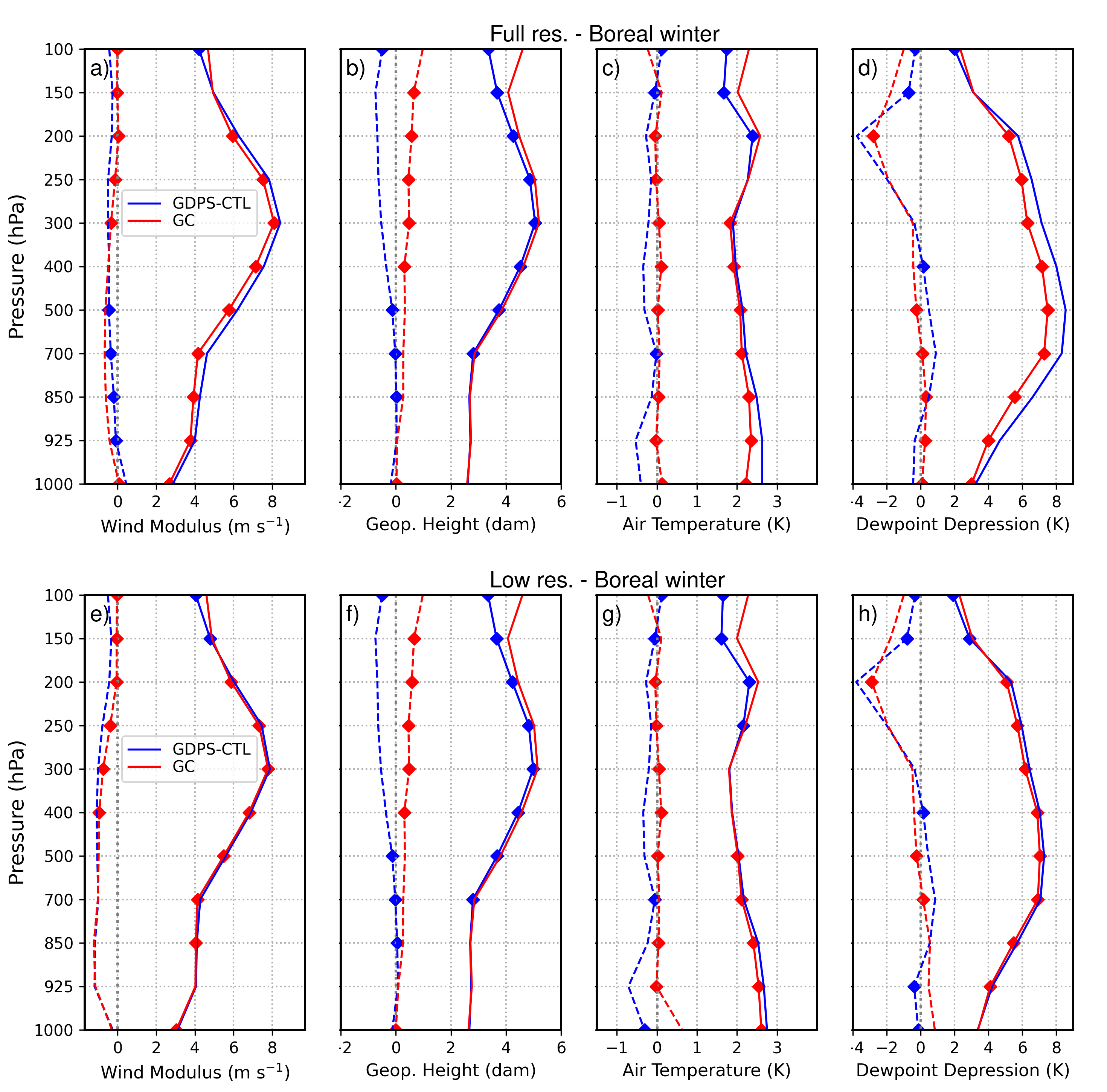}\\
\caption{Verification for 120-hr forecasts from GraphCast (red) and GDPS-CTL (blue) at their full (top row: a--d) and low (bottom row: e--h) resolutions against global radiosonde observations, averaged over 60 cases in boreal winter. The error standard deviation (STDE; solid lines) and bias (mean forecast-minus-observation; dashed lines) are shown for (a,e) wind modulus, (b,f) geopotential heights, (c,g) temperature, and (d,h) dewpoint depression. Red and blue markers denote statistically significant differences in favor of GraphCast and GDPS-CTL, respectively. Significance is computed using the $t$-test for bias and the $F$-test for STDE. No marker at a pressure level implies that the null hypothesis, stating that the statistics of the two samples are the same, cannot be rejected based on the 90th percentile. A low pass filter (appendix B) is applied on both GraphCast and GDPS-CTL to obtain the low-resolution fields (e--h).}
\label{f_arcad_1}
\end{center}
\end{figure}

\subsubsection{At full resolution}
Comparing the forecasts at their native resolutions against radiosonde observations reveals that GraphCast forecasts are generally closer to the observations than GDPS-CTL in the troposphere. Figures \ref{f_arcad_1}a--d show the corresponding results at day 5. While GraphCast shows clear improvements in temperature biases (dashed lines), it suffers from some noticeable deceleration within the troposphere, leading to deteriorations in wind modulus biases. In terms of error standard deviation (STDE, solid lines), however, GraphCast clearly shows large improvements over GDPS-CTL for wind modulus and dewpoint depression, although suffers from some deterioration in the mid-to-upper troposphere geopotential height. Above 200 hPa, GraphCast suffers from increased errors, with biases and STDE being comparable or worse than those in GDPS-CTL.

To summarize the scores at different lead times, variables, and vertical levels, a single forecast quality index based on the change in root-mean-square error (RMSE) is defined as

\begin{equation}
 \label{e_rmse}
 f_{RMSE}=\left[\frac{RMSE(x_1)}{RMSE(x_2)}-1\right]\times 100\%
\end{equation}

\noindent where $x_1$ and $x_2$ represent GDPS-CTL and GraphCast, respectively. Positive values of $f_{RMSE}$ indicate improved forecasts from GraphCast, while negative values imply degradations. A $f_{RMSE}$ was computed every 24 hr for each selected variable and level, with the results vertically averaged over three layers: low level (from 1000 to 850 hPa), mid level (from 700 to 300 hPa), and upper level (from 250 hPa to 100 hPa).

\begin{table}[h]
\caption{Latitude and longitude bounds for the 6 subdomains considered in Figures \ref{f_heatmap_1}, \ref{f_heatmap_2} and \ref{f_heatmap_3}.}\label{t_1}
\begin{center}
\begin{tabular}{cccccrrcrc}
\topline
 Domain & Southern Bound & Northern Bound & Western Bound & Eastern Bound\\
\midline
 Northern Hemisphere  & 20N & 90N & -    & - \\
 Southern Hemisphere  & 90S & 20S & -    & - \\
 Tropics              & 20S & 20N & -    & - \\
 North America        & 25N & 85N & 170W & 40W \\
 Europe               & 25N & 70N & 10W  & 28E \\
 Asia                 & 25N & 60N & 65E  & 145E \\
\botline
\end{tabular}
\end{center}
\end{table}

\begin{figure}
\begin{center}
\noindent\includegraphics[width=38pc,angle=0]{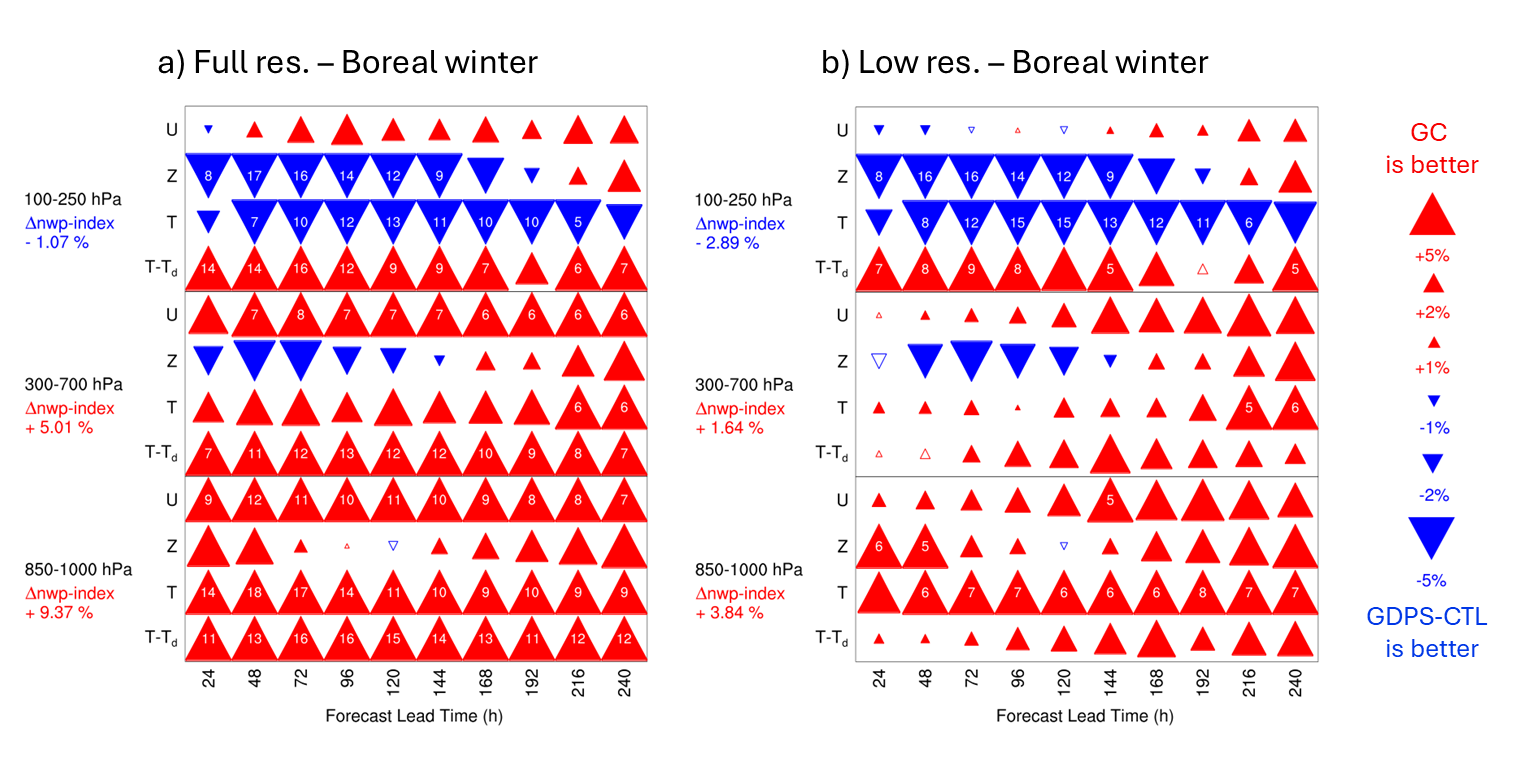}\\
\caption{Changes in the forecast quality index (Eq. \ref{e_rmse}) against global radiosonde observations, averaged over 3 layers and 60 cases during boreal winter 2022, corresponding to forecasts at (a) full resolution and (b) low resolution. Upward-pointing red triangles indicate a reduction of RMSE by GraphCast with respect to GDPS-CTL, whereas downward-pointing blue triangles indicate the opposite. The size of the triangles varies linearly up to a value of 5.0\%. When the index is greater than this threshold, the size is kept constant, but the rounded values are shown. Triangles are color-filled if the significances computed using an $F$-test exceeds the 90th percentile. The values on the left-hand side show the averaged indices over all the lead times and all the variables for each layer.}
\label{f_arcad_scorecard_1}
\end{center}
\end{figure}

Results for boreal winter show that $f_{RMSE}$ is quite constant with lead times for most variables and layers (Fig. \ref{f_arcad_scorecard_1}a). The largest improvements from GraphCast are obtained in the low-level layer for all variables and decrease with altitude, turning into deteriorations in the upper-level layer for geopotential heights and temperatures. GraphCast geopotential heights are also degraded with respect to GDPS-CTL in the mid-level layer at short lead times, which is consistent with the error profile presented in Fig. \ref{f_arcad_1}b. Results for boreal summer are roughly similar (not shown). The pressure weighting approach adopted for the training of GraphCast (see Fig. 6 in \citealt{lsw22}) is likely responsible for this vertical variation in performances. Systematic differences in the weakly constrained upper-level layer between ECMWF (used for fine-tuning) and GDPS analyses (used for initializing GraphCast) -- e.g., different bias correction approaches for radiances, different data assimilated, distinct biases in the IFS and GEM model, etc. -- could also explain some of the degradations observed in the upper-level layer. However, GraphCast initialized with ECMWF-ERA5 has also been shown to increase stratospheric RMSE with respect to ECMWF analyses (see, e.g., Fig. 19 in \citealt{lsw22}).

\begin{figure}
\begin{center}
\noindent\includegraphics[width=28pc,angle=0]{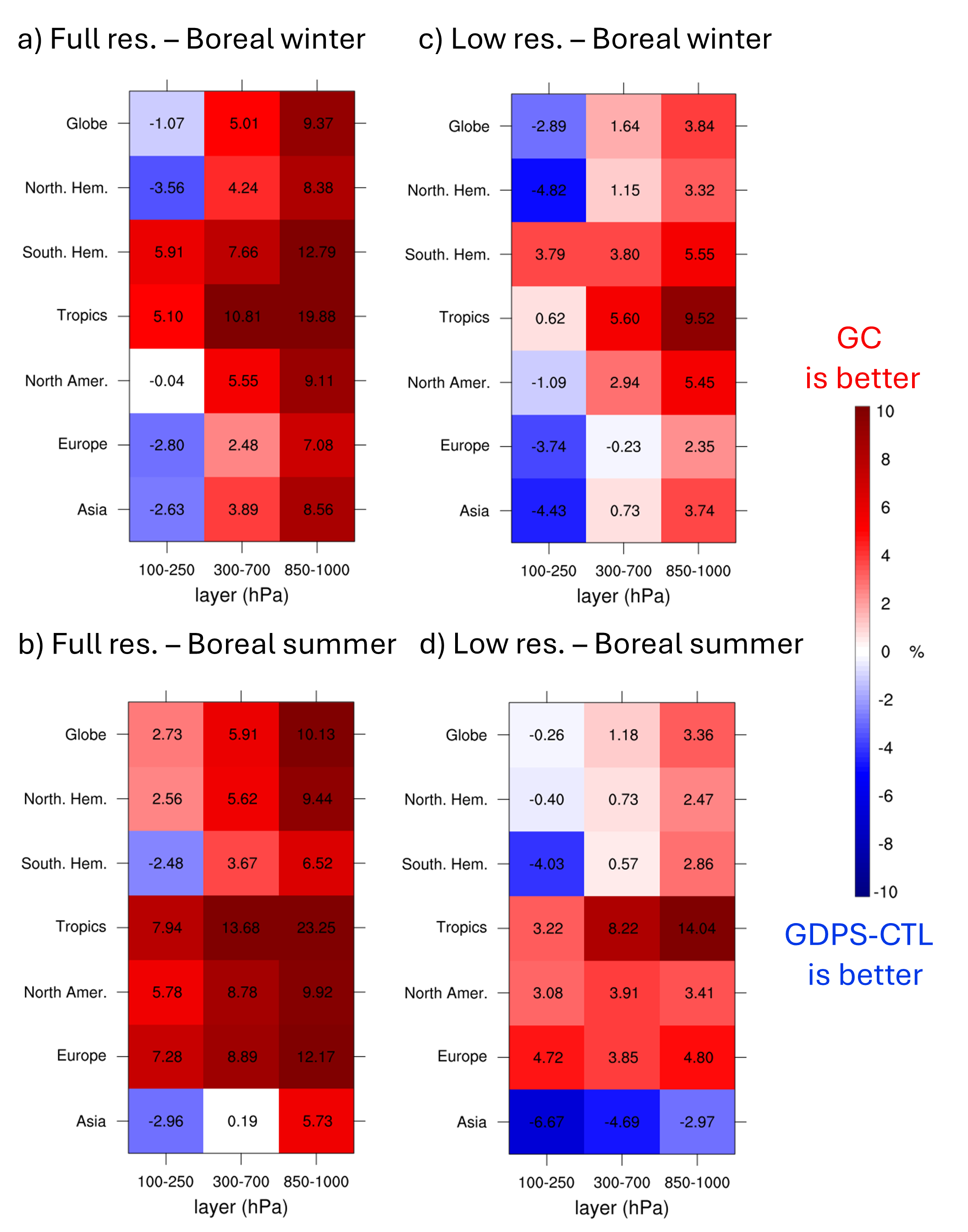}\\
\caption{Heatmap of the changes in the forecast quality index (Eq. \ref{e_rmse}) against global radiosonde observations for various geographical domains and atmospheric layers averaged over all lead times for (a, c) 60 cases of boreal winter 2022, and (b, d) 62 cases of boreal summer 2022. Verifications are presented for forecasts (a, b) at their full resolution, and (c, d) after applying a low pass filter (appendix B) on both GraphCast and GDPS-CTL. Red shadings indicate a reduction of the RMSE by GraphCast with respect to GDPS-CTL, while blue shadings indicate the opposite.}
\label{f_heatmap_1}
\end{center}
\end{figure}

Regional variations in relative forecast performance can be examined by computing $f_{RMSE}$ on the six subdomains listed in Table 1. In low- and mid-level layers, GraphCast forecasts show smaller RMSE in every subdomain in both seasons (Figs. \ref{f_heatmap_1}a and b). At upper levels, GDPS-CTL performs better during each hemisphere’s respective winter and over Asia during boreal summer. However, GraphCast's poor performance during winter over the Northern Hemisphere is primarily originating from Europe and Asia. The largest improvements from GraphCast in the mid- and upper-level layers are found over the Tropics, as  reported by \cite{lsw22} (see their Fig. 19). This consistency in GraphCast's regional performance with respect to its initialization with different analyses (ECMWF and GDPS analyses) is reassuring.

\subsubsection{At low resolution}
At low resolution, GraphCast forecast improvements over GDPS-CTL are reduced for all variables, levels, lead times, seasons, and geographical domains (Figs. \ref{f_arcad_1}e--h, \ref{f_arcad_scorecard_1}b, and \ref{f_heatmap_1}c,d). The change is particularly striking in the mid-level layer over the first 72 hr (compare Figs. \ref{f_arcad_scorecard_1}a and b), with considerably smaller GraphCast forecast improvements than in the full-resolution case. The forecast improvements in low- and full-resolution context, however, become more aligned around 120 hr (Fig. \ref{f_arcad_scorecard_1}b). This stands in contrast with a relatively steady improvement at all lead times found at full resolution (Fig. \ref{f_arcad_scorecard_1}a), implying that these apparent short-range GraphCast forecast improvements, with respect to GDPS-CTL, originate primarily from partial smoothing of the less-accurately predicted fine scales that reduces the double penalty. At low resolution, as expected, both GraphCast and GDPS-CTL show a worsening wind speed bias resulting from the suppression of extremes (Fig. \ref{f_arcad_1}e); however, there is no statistically significant difference between their predictions. This suggests that the degradation observed in GraphCast at full resolution, relative to to GDPS-CTL (Fig. \ref{f_arcad_1}a), is primarily due to fine-scale smoothing.

After filtering, regional deteriorations in GraphCast predictions are accentuated and regional improvements are significantly reduced, with the average reduction factor for $f_{RMSE}$ from full- to low-resolution verification being roughly between 3 to 4\% lower than the values obtained at full resolution (c.f. the columns in Fig. \ref{f_heatmap_1}). This further confirms that the lower effective resolution of GraphCast provides a significant advantage over the GDPS-CTL. Nevertheless, over most regions, and for both seasonal periods, the large scales in GraphCast forecasts are still, on average, considerably closer to tropospheric radiosonde observations than those from GDPS-CTL, implying that they contain useful large-scale information that could substantially improve guidance if successfully incorporated into a hybrid NWP-AI system.

\section{Spectral nudging in GEM}
\label{s_spn}
\subsection{The concept}
Spectral nudging works by directing the model-predicted atmospheric large scales toward a more skillful reference \citep{slf00,hsy14}. This approach fundamentally differs from grid nudging or indiscriminate nudging \citep{lth12}, as it only targets a predefined range of scales. Even though an NWP model may employ grid-based spatial discretizations, the nudging increments at a given model level are computed by decomposing the model predictions and the reference fields in a spectral space, retaining only the target scales. Hence, this technique is referred to as ``spectral nudging''.

In limited-area configuration, such as regional climate models, the nudging reference is usually derived from the same data that provides the lateral boundary conditions (LBCs). For extended-range integrations -- spanning weeks to months or even years -- the LBCs  alone are insufficient to prevent the model-predicted large scales from drifting significantly. Consequently, spectral nudging becomes essential \citep{lle09}. For multi-month kilometer-scale downscaling applications, the evolving surface fields may also exhibit unacceptable deviations, necessitating nudging toward a reliable reference, as discussed by \cite{hsy14}.

\subsection{Implementation in GEM}
With the split-type dynamics-physics coupling in GEM, the adiabatic dynamical core first solves the prognostic dynamical equations to obtain an intermediate state of the atmosphere. In the absence of spectral nudging, this interim dynamics solution serves as the input for the physical parameterization schemes, which compute the physics-induced tendencies. These tendencies are then coupled with the dynamics solution to obtain the complete model solution for a given time step. However, when spectral nudging is applied, the large scales in the solutions for a selected set of prognostic variables from GEM dynamics are first nudged toward the reference (here, from GC). Subsequently, the nudged solutions are fed to the physics schemes to compute the physics-induced tendencies before the eventual dynamics-physics coupling.

In mathematical form, the nudging step at a given model vertical level, i.e., a constant-$\zeta$ surface, can be expressed as

\begin{equation}
\label{e_nudge}
  F_{nudge} = F_{GEM} + \omega \left[F_{GC}-F_{GEM}\right]_{LS},
\end{equation}

\noindent where $F_{GEM}$ is a prognostic variable predicted by GEM dynamics, $F_{GC}$ is the corresponding prediction from GC vertically interpolated to the $\zeta$ level valid at the same time, $\omega$ is the nudging relaxation factor such that $(0 \le\omega\le 1)$, and $F_{nudge}$ is the nudged solution. The subscript $LS$ in the above equation refers to some user-defined large scales targeted by the nudging mechanism, and is discussed later in more detail. Through simple rearrangement of the terms on the right-hand-side of Eq.\ref{e_nudge}, it can be shown that the large scales in $F_{nudge}$ are indeed a weighted average of those from $F_{GEM}$ and $F_{GC}$ (with $\omega$ being the weighting parameter), whereas the small scales from $F_{GEM}$ are retained entirely in $F_{nudge}$.

As shown by \cite{hsy14}, $\omega$ can vary with model vertical levels and in time, and is given by

\begin{equation}
\label{e_omega}
  \omega=\frac{\beta(\zeta)}{\tau(t)}\Delta t,
\end{equation}

\noindent where $\beta(\zeta)$ defines the nudging vertical profile with respect to the $\zeta$-coordinate, $\tau(t)$ represents the nudging relaxation time scale , and $\Delta t$ denotes the model time-step length. A vertically-variable $\beta$ allows for different nudging strengths at different model levels. Conversely, a time-dependent $\tau$ permits changing nudging strengths with time. In general, a large value of $\tau$ implies weak nudging, and vice versa.

To nudge a prognostic variable $F$ at any model level, it is first essential to spectrally decompose $(F_{GC}-F_{GEM})$ so that a filter can be applied in the spectral space to retain only the desirable scales. For global simulations, such a filter should ideally employ a spherical harmonics-based spectral decomposition. However, this would require multiple transformations of model solutions between the model's overlapping limited-area Yin-Yang grids and an intermediate global Gaussian grid, resulting in significant increase in the computational cost. As the present study is primarily of a proof-of-concept nature, the spectral filter for nudging was instead chosen to be based on discrete cosine transform (DCT) \citep{dcl02}, and the filtering is performed separately on the Yin and Yang grids. The DCT-based spectral filter employs two cut-off wavelengths for large and small scales, denoted by $\lambda_{LS}$ and $\lambda_{SS}$, respectively. Scales larger than $\lambda_{LS}$ are fully retained by the filter, whereas scales smaller than $\lambda_{SS}$ are entirely removed, with a partial filtering of the scales in between (appendix C). This partial filtering, often referred to as soft cut-off \citep{hsy14}, helps to minimize Gibbs oscillations \citep{sho84}. The values of $\lambda_{LS}$ and $\lambda_{SS}$ are selected to produce a qualitatively acceptable response of the filter when tested over a global Gaussian grid of equivalent resolution (appendix C). Overall, the evaluation of $\left[F_{GC}-F_{GEM}\right]_{LS}$ for the purpose of spectral nudging involves computing the spectral coefficients of $(F_{GC}-F_{GEM})$ using a DCT, followed by the application of the spectral filter described in appendix C. An inverse DCT is then applied to obtain the large-scale differences in physical space.

\subsection{Optimal configuration}
A series of systematic sensitivity experiments was carried out to identify an optimal spectral nudging configuration for the hybrid GDPS forecasts, referred to as GDPS-SN. Note that spectral nudging is currently only applied within the forecasting component of GDPS, i.e., GDPS-CTL, GDPS-SN, and GraphCast are all initialized with the same analyses.

Some key aspects of the optimal nudging configuration are discussed below.

\subsubsection{Nudged variables}
Nudging is only applied to the $u$-$v$ components of wind and virtual temperature. While nudging specific humidity can improve the global bias of temperature and humidity in the boundary layer, the computational cost (see section \ref{s_eva}\ref{s_eva_cc}) outweighs the benefits. Moreover, sensitivity experiments have revealed a negative impact of specific humidity nudging on tropical cyclone intensity (not shown), making it less desirable.

\subsubsection{Nudging vertical profile}
\begin{figure}
\begin{center}
\noindent\includegraphics[width=17pc,angle=0]{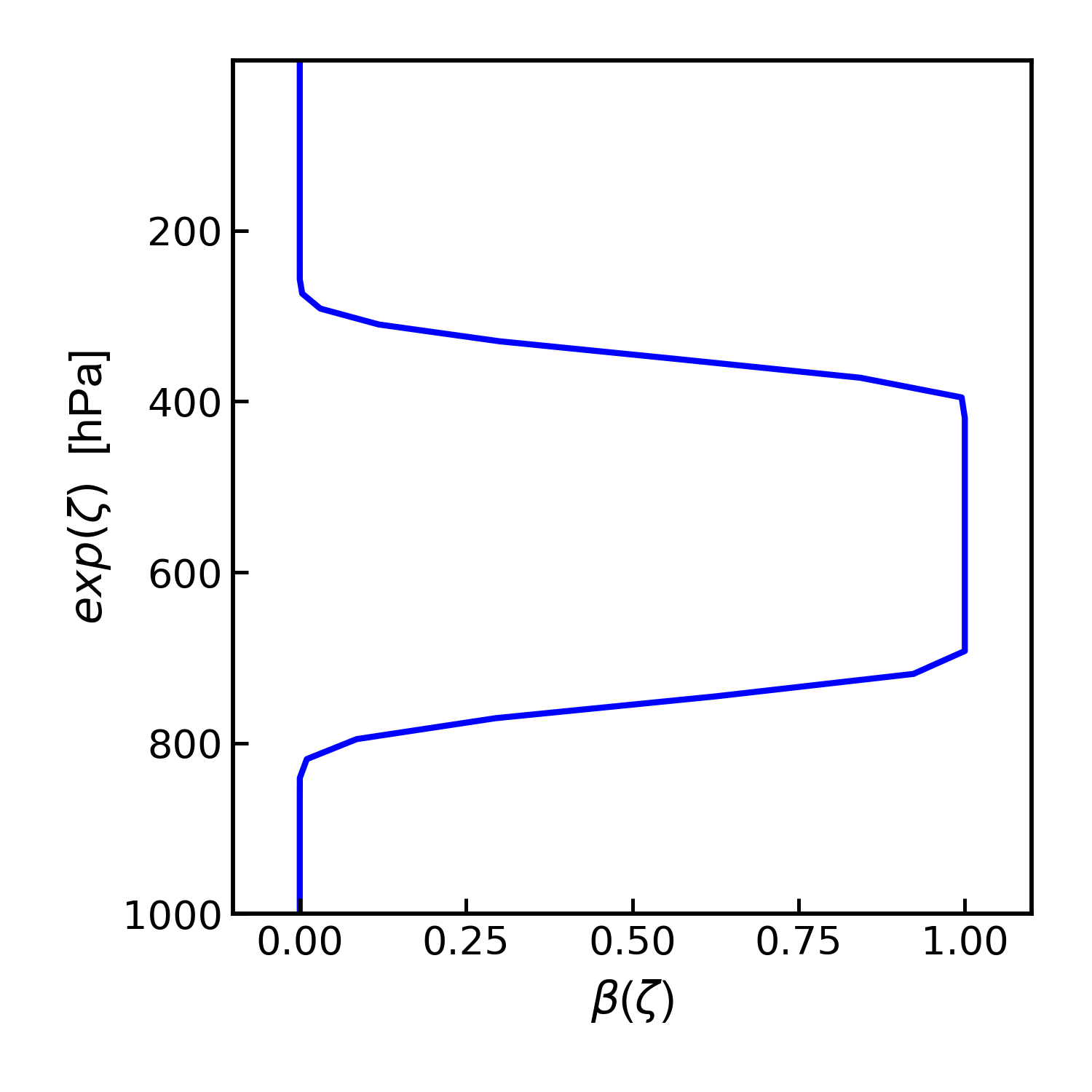}\\
\caption{Nudging vertical profile for the presented optimal configuration, where $0\leq\beta(\zeta)\leq1$ with $\zeta$ being the log-hydrostatic-pressure-type terrain-following vertical coordinate in the operational GEM model. The value of $\beta$ is set to zero when $\exp(\zeta)>$850 hPa and $\exp(\zeta)<$250 hPa. Its value is set to 1 for 400 hPa$\leq\exp(\zeta)\leq$700 hPa. Transition of $\beta$ from 0 to 1 (and vice versa) follows a cosine-squared profile similar to the one presented by \cite{hsy14}.}
\label{f_np}
\end{center}
\end{figure}

Although different vertical profiles, defined by $\beta(\zeta)$, have been explored, a plateau-shaped profile, as shown in Fig. \ref{f_np}, is found to yield the best results. This profile involves no nudging in the boundary layer (below 850 hPa) and the stratosphere (above 250 hPa). Relatively weaker performance of GC in the stratosphere -- as mentioned earlier -- is the reason for no stratospheric nudging. Although nudging is generally avoided in the boundary layer to allow uninhibited generation of fine scales (e.g., \citealt{sff17}), other studies have found the potential for significant near-surface skill improvement with nudging towards an accurate reference in the boundary layer (e.g., \citealt{hsy14}). However, in this study, nudging is avoided in the boundary layer for different reasons. First, the 13-level version of GC (used here) does not have sufficient vertical resolution in the boundary layer to provide a useful reference. Even in the absence of this issue, inconsistencies between the surface forcing in GDPS and GC training data (ERA5) may lead to potential negative impacts over regions with complex terrain. In addition, differences in mean boundary-layer state (moisture in particular) may result in adverse reactions from the physical parameterizations in GEM. Therefore, nudging in the boundary layer is not a viable option at this stage.

\subsubsection{Nudging length scales}
It was shown earlier in section \ref{s_rp} that the fine-scales in GC suffer from considerable smoothing, worsening at longer lead times (see Figs. \ref{f_tot_spc_1} and \ref{f_var_coh_1}). The implementation of spectral nudging in GEM allows for time-varying cut-off scales for nudging. However, improvements with such an approach over stationary cut-off scales are found to be negligible. Therefore, based on the comparison of spectra between GDPS and GC, the DCT-based filter for the optimal configuration utilized in this study is configured with $\lambda_{LS}$=2750 km and $\lambda_{SS}$=2250 km. This implies that scales larger than 2750 km are entirely retained, while those smaller than 2250 km are completely eliminated. The scales in between are subjected to partial filtering (see Fig. \ref{f_a1}). Over a global Gaussian grid of comparable resolution, the filter response is approximately equivalent to a soft cut-off between 5500-km and 2000-km scales (i.e., $\lambda_{LS}$=5500 km and $\lambda_{SS}$=2000 km), which provides a more relevant context for interpreting the results. This choice of nudging length scales corresponds to the effective resolution of GC beyond day 3 (see Figs. \ref{f_tot_spc_1} and \ref{f_var_coh_1}).

\subsubsection{Nudging relaxation time}
The nudging relaxation time, $\tau$, is analogous to the $e$-folding time as it determines the time-rate of decay in the large-scale differences between the model and the driving data in the absence of any other process. Arguably, $\tau$ is one of the most important configurable parameters for spectral nudging. However, selecting an optimal $\tau$ is not straightforward. First, it is important to note that GC inferences are only available every 6 hours. During model integration, when a valid GC inference is not available at certain time steps, an estimate is approximated through linear interpolation between two consecutive GC inferences around the time of interest. In this regard, \cite{odd12} have argued that $\tau$ should not be less than $\tau_a$, which denotes the time interval between consecutive driving fields (here, GC inferences). \cite{hsy14} showed that, with $\tau<\tau_a$, the impact of time interpolation error on the nudged fields may not be negligible. As a result, $\tau_a$ may be considered the lower bound for $\tau$.

It is also crucial to ascertain an acceptable upper bound for $\tau$. Sensitivity experiments have revealed that too large of a nudging relaxation time ($\tau>24$ hr) can result in excessive undesirable smoothing in the nudged fields around the cut-off scales developing at day 3 or 4. Eventually, this smoothing affects all scales by day 10 of the forecast (not shown). This increased smoothing with overly large $\tau$, is caused by the averaging effect resulting from blending two forecasts, with $\tau=48$ hr maximizing the effect.

Strong nudging with $\tau=\tau_a$ is also found to result in erroneous evolution of nudged fields emanating from inconsistencies between GEM and GC over complex terrain, as well as error originating from vertically interpolating GC to GEM levels. Therefore, based on the insights derived from the sensitivity tests, a nudging relaxation time of 12 hr is chosen to be optimal. An important point to consider in this context is that although GC inferences become increasingly inaccurate with respect to GDPS analyses at longer lead times, they still maintain higher spectral coherence (compared to GDPS-CTL) for the largest scales (see Fig. \ref{f_var_coh_1}c,d). Therefore, a constant value of $\tau$ during the entire integration is justified.

\section{Impact of large-scale spectral nudging}
\label{s_eva}

\subsection{Verification with global power spectra}
\begin{figure}
\begin{center}
\noindent\includegraphics[width=38pc,angle=0]{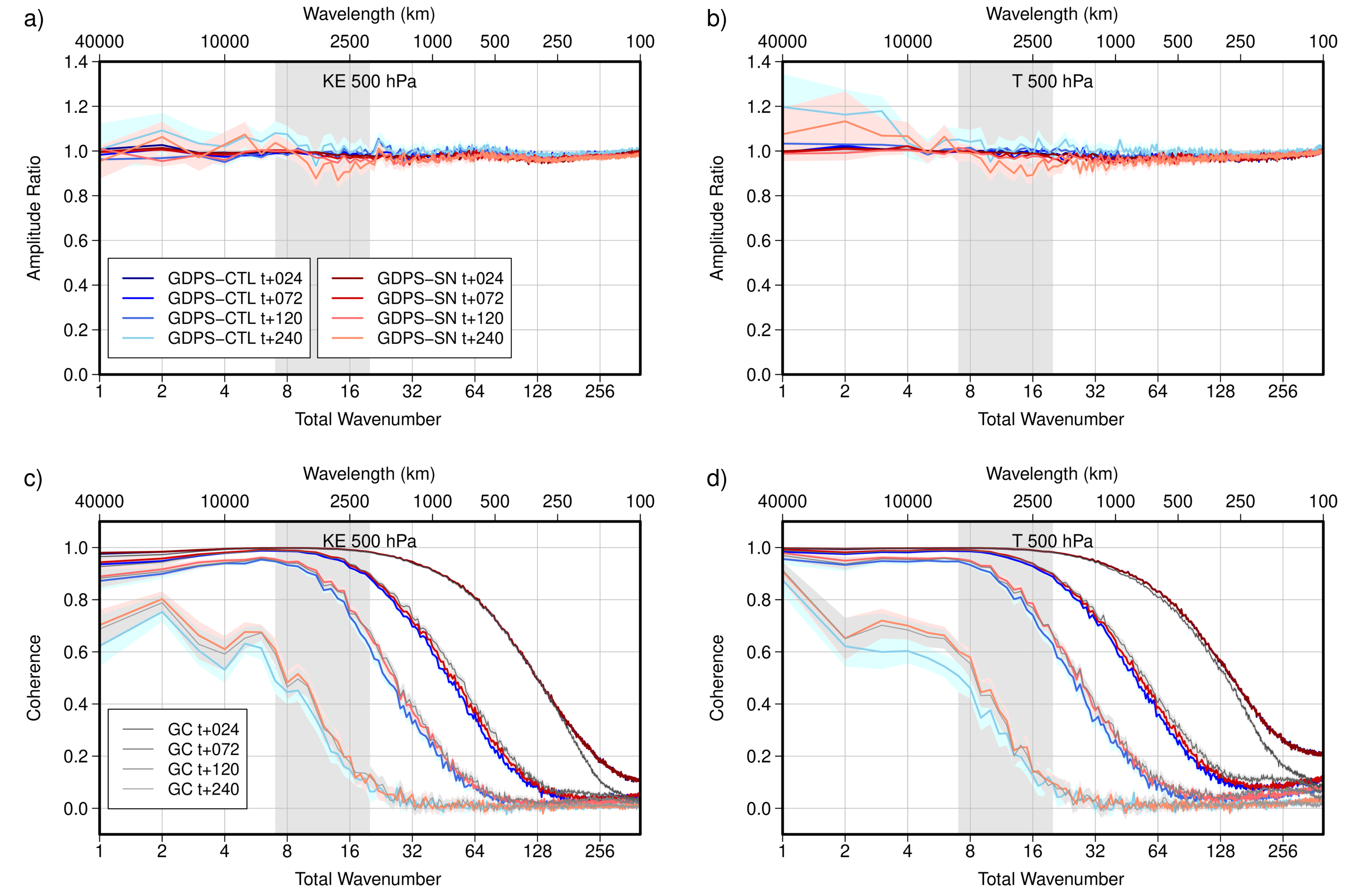}\\
\caption{Same as in Fig. \ref{f_var_coh_1}, but for GDPS-SN and GDPS-CTL. Additionally, for convenience of comparison, GraphCast has been added in Figs. \ref{f_var_coh_2}c and d. The small- and large-scale cut-offs for the DCT-based filter over a global Gaussian grid (appendix C) are approximately 2000 km and 5500 km, respectively, as indicated with the shaded area.}
\label{f_var_coh_2}
\end{center}
\end{figure}

The power spectra of the GDPS-SN forecast variables are indistinguishable from GDPS-CTL (not shown), which essentially  suggests that the new hybrid system can resolve the full range of scales present in the GDPS-CTL without significant smoothing. The resulting spectral amplitude ratio of transient-eddy anomalies, $\gamma$ (Eq. \ref{e_var}), for GDPS-SN also has values very close to the desired value of 1 for all prognostic variables, lead times, and spatial scales (Figs. \ref{f_var_coh_2}a, b). The only exception is a 10\% reduction of $\gamma$ for scales between 2000 km and 4000 km at lead times approaching day 10 (wavenumbers 10--20 in Figs. \ref{f_var_coh_2}a, b). This reduction is explained by GraphCast having some variance deficiency at the corresponding range of scales (2000--2750 km) for longer lead times (beyond day 4; Figs. \ref{f_var_coh_1}a, b). Sensitivity tests with time-varying cut-off length scales ($\lambda_{LS}$ and $\lambda_{SS}$) that more strictly follow GraphCast’s time-evolving effective resolution showed that this issue could be avoided, but at the expense of significantly reduced GDPS-SN forecast improvements with respect to the GDPS-CTL. This is because larger cut-offs imply a weaker GraphCast-based constraint at synoptic scales in the hybrid system.

Comparison of the spectral coherence (Eq. \ref{e_coh}) between the forecasts and GDPS analysis shows improved skill in GDPS-SN over GDPS-CTL (Figs. \ref{f_var_coh_2}c, d). The improvements are somewhat smaller than those seen with GraphCast at shorter lead times (up to day 3), but become comparable thereafter (Figs \ref{f_var_coh_2}c, d). The scales relevant for short-range improvements are smaller than those targeted by the chosen spectral nudging configuration. Inclusion of finer scales for nudging would improve short-range coherence, but it would result in problematic variance deficiency at longer leads.

\begin{figure}
\begin{center}
\noindent\includegraphics[width=38pc,angle=0]{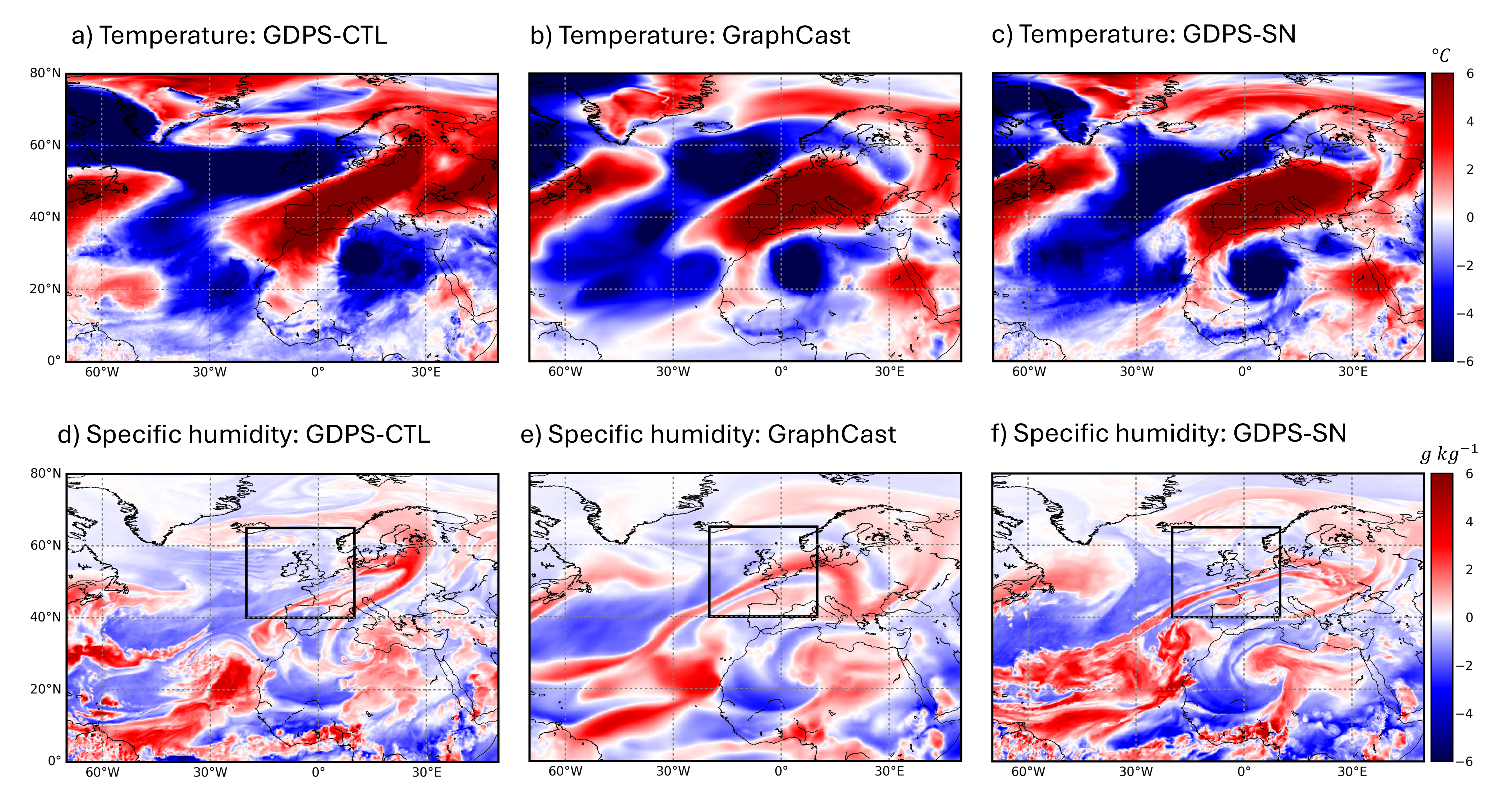}\\
\caption{10-day forecast of 700-hPa transient-eddy anomalies for (a--c) temperature ($\degree$C), and (d--f) specific humidity (g kg$^{-1}$), valid at 0000 UTC, 18 February 2022 for GDPS-CTL (left), GraphCast (middle), and GDPS-SN (right). The black rectangles in Figs. \ref{f_smooth} d-f indicate the region that is later used for presenting results in Fig. \ref{f_eunice}.}
\label{f_smooth}
\end{center}
\end{figure}

The impact of spectral nudging in the physical space is shown using maps of transient-eddy anomalies of temperature and specific humidity in Fig. \ref{f_smooth}. Nudging ensures that the large-scale temperature anomalies closely resemble GraphCast (Figs. \ref{f_smooth}a--c). Even for specific humidity, a variable which is not directly constrained by nudging, the large scales are found to be more aligned with GraphCast than with GDPS-CTL (Figs. \ref{f_smooth}d--f). This reinforces the findings of previous studies on spectral nudging \citep{hsy14}. Regarding fine scales, Fig. \ref{f_smooth} shows that GDPS-SN has similar level of variability as GDPS-CTL in contrast to the heavily smoothed GraphCast fields.

\begin{figure}
\begin{center}
\noindent\includegraphics[width=38pc,angle=0]{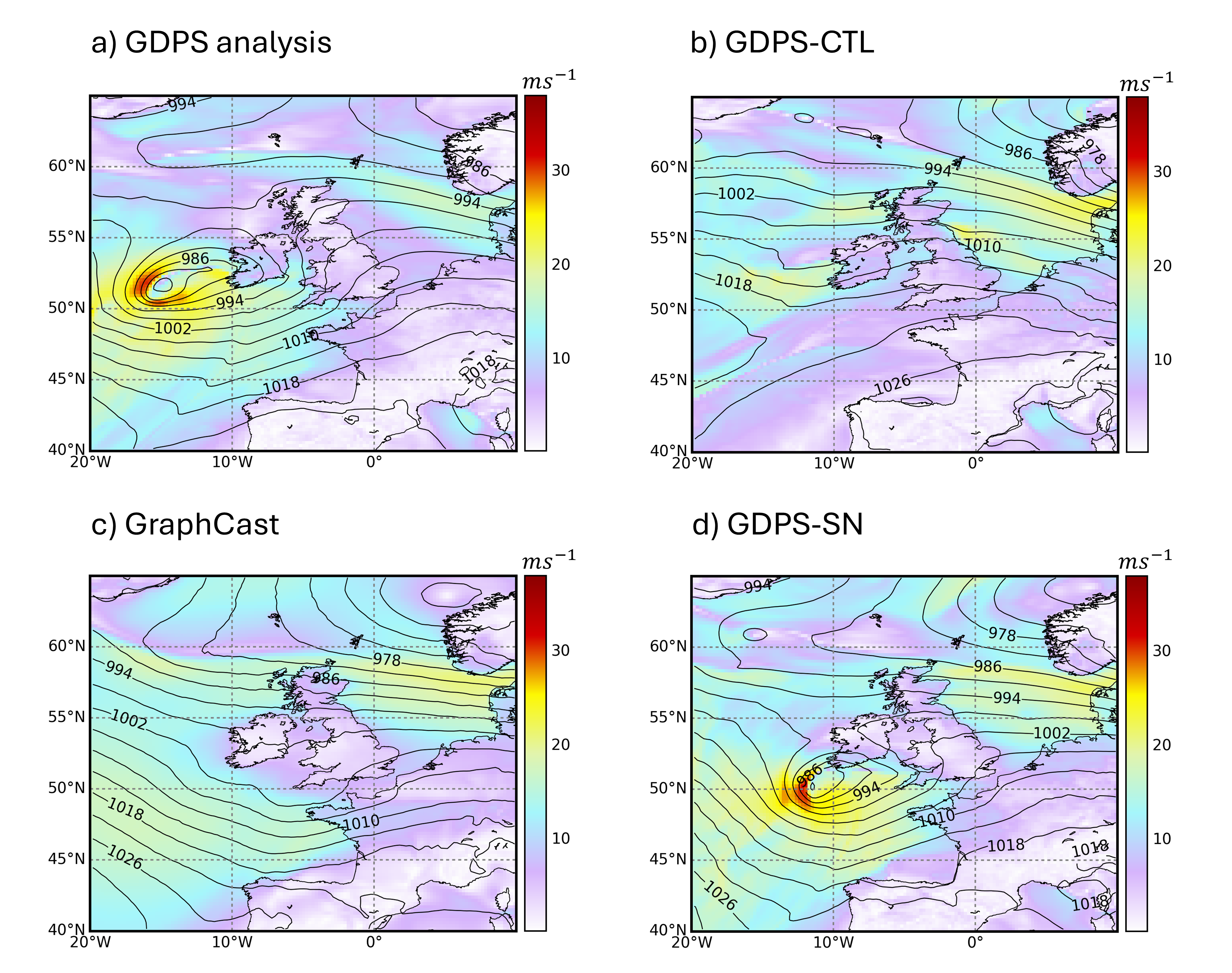}\\
\caption{Maps of mean sea-level pressure (hPa) and 10-m wind speed (m s$^{-1}$), valid at 0000 UTC, 18 February 2022 from: (a) GDPS analyses, and 5.5-day forecasts from (b) GDPS-CTL, (c) GraphCast, and (d) GDPS-SN. Results are presented over a domain identified with a black rectangle in Fig. \ref{f_smooth}d--f.}
\label{f_eunice}
\end{center}
\end{figure}

An illustrative example of the value added by the proposed hybrid system is presented in Fig. \ref{f_eunice}, which shows the 5.5-day forecasts by GDPS-CTL, GraphCast, and GDPS-SN, as well as GDPS analysis, valid at 0000 UTC on 18 February 2022. GDPS analysis shows the presence of the extratropical winter storm Eunice, which is also predicted by GDPS-SN, whereas both GDPS-CTL and GraphCast fail to predict the storm for this lead time. It is important to note that all three models are capable of predicting the storm at the next initialization time (0000 UTC on 14 February 2022; not shown). These results imply that, for a lead time of 5.5 days, although GraphCast may predict large scales that are favorable for the formation of Eunice, it fails to develop the storm, presumably due to excessive smoothing at scales smaller than 2000 km. Since GDPS-SN does not suffer from similar smoothing, it can effectively leverage GraphCast's more accurate large-scale information, resulting in a significant gain in predictability for this specific event, compared to both GraphCast and GDPS-CTL. More importantly, these results demonstrate that large-scale nudging of GDPS-SN toward GraphCast can lead to mesoscale features that are substantially different from both GDPS-CTL and GraphCast. In other words, the spectral nudging-based hybrid system has the potential to add substantial value beyond what is attainable through any offline post-processing method that combines NWP and AI models.

\subsection{Verification against radiosondes}
\begin{figure}
\begin{center}
\noindent\includegraphics[width=38pc,angle=0]{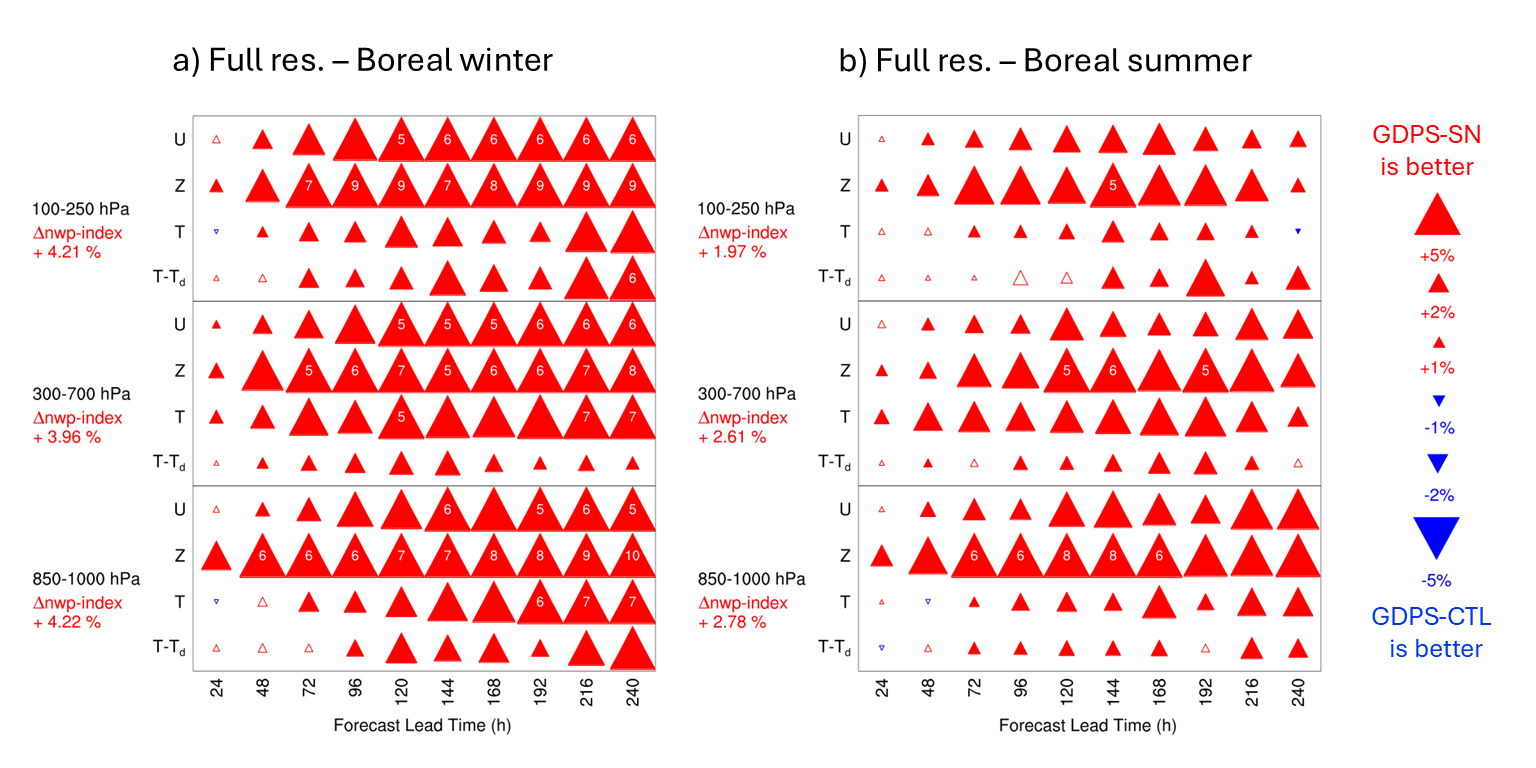}\\
\caption{Same as in Fig. \ref{f_arcad_scorecard_1}, except that upward-pointing red triangles indicate a reduction of RMSE by GDPS-SN with respect to GDPS-CTL, whereas downward-pointing blue triangles indicate the opposite. Only verification using forecasts at full resolution is shown.}
\label{f_arcad_scorecard_2}
\end{center}
\end{figure}

Comparison of forecasts at full resolution against the global radiosonde observations clearly indicates that GDPS-SN forecasts are closer to the observations than GDPS-CTL forecasts across variables, lead times, and levels for both seasons (Fig. \ref{f_arcad_scorecard_2}). The improvements grow gradually over the first 72 hr and are remarkably constant with altitude, which contrasts with the strong vertical variations observed with GraphCast in Section \ref{s_rp}. This demonstrates that large-scale spectral nudging, despite being only applied to free-tropospheric winds and temperature, is sufficient to impart considerable improvements to the unconstrained upper- and lower-level layers. Compared to boreal winter (Fig. \ref{f_arcad_scorecard_2}a), the improvements from GDPS-SN are generally reduced during boreal summer (Fig. \ref{f_arcad_scorecard_2}b), which is consistent with the relative performance of the component systems (not shown).

\begin{figure}
\begin{center}
\noindent\includegraphics[width=28pc,angle=0]{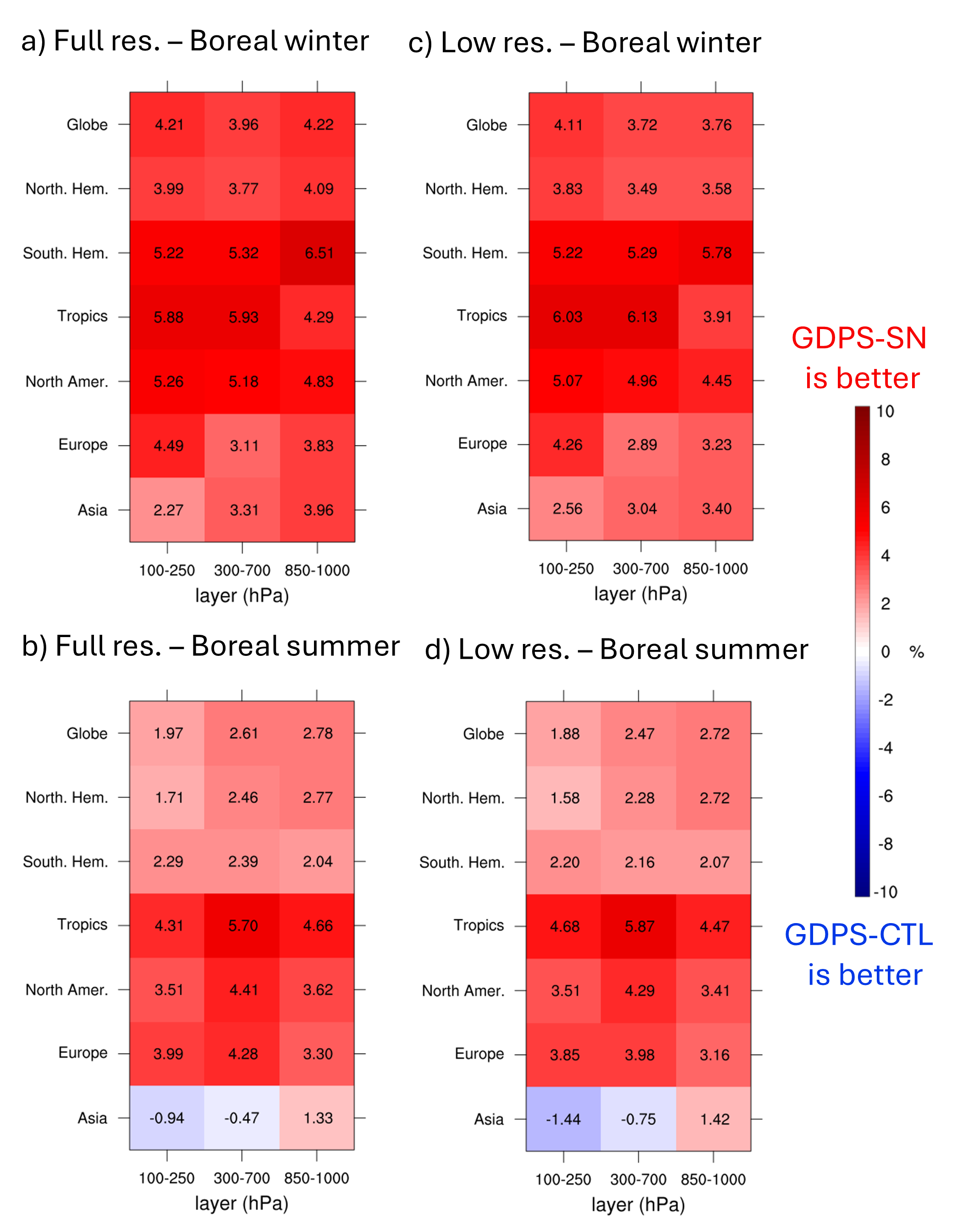}\\
\caption{Same as in Fig. \ref{f_heatmap_1}, except that red shadings indicate a reduction of RMSE by GDPS-SN with respect to GDPS-CTL, while blue shadings indicate the opposite.}
\label{f_heatmap_2}
\end{center}
\end{figure}

Regionally, GDPS-SN has smaller RMSE than GDPS-CTL in every subdomain except over Asia in the mid- and upper-levels during boreal summer (Figs. \ref{f_heatmap_2}b and d), which is consistent with the degradations noted for GraphCast large scales in this region (Fig. \ref{f_heatmap_1}d). Unlike GraphCast, the forecast improvements of GDPS-SN over GDPS-CTL, when measured against radiosonde observations, are comparable at both full resolution (left column of Fig. \ref{f_heatmap_2}) and for large scales only (right column of Fig. \ref{f_heatmap_2}, compare Fig. \ref{f_heatmap_1}), showing only a small degradation under the low-pass filter. This implies that the fine scales predicted by GDPS-SN benefit from fractional improvements at large scales, which is supported by the improvements in fine-scale spectral coherence with GDPS-SN over GDPS-CTL, particularly beyond day 3 (Fig. \ref{f_var_coh_2}c-d).

\begin{figure}
\begin{center}
\noindent\includegraphics[width=38pc,angle=0]{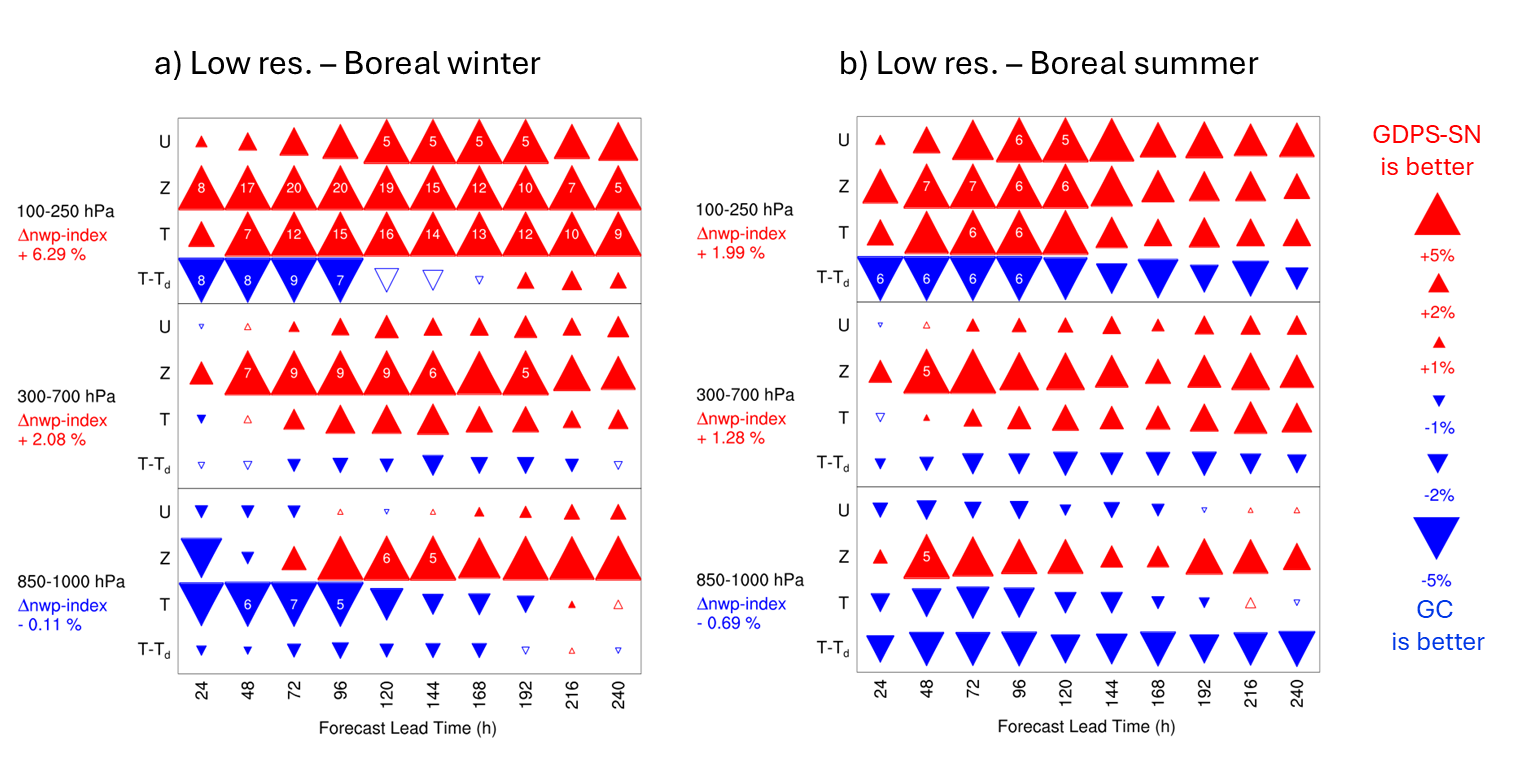}\\
\caption{Same as in Fig. \ref{f_arcad_scorecard_1}, except that upward-pointing red triangles indicate a reduction of RMSE by GDPS-SN with respect to GraphCast, whereas downward-pointing blue triangles indicate the opposite. Only verification using forecasts at low resolution is shown, but for both seasons.}
\label{f_arcad_scorecard_3}
\end{center}
\end{figure}

Forecasts from GDPS-SN must also be compared to those from GraphCast to ensure that the skill of the AI-based system is not lost during hybridization. For the sake of fair comparison, this is only done at low resolution (section \ref{s_rp}b2). Predictions from GDPS-SN are better than GraphCast in the upper- and mid-level layers, except for dewpoint depression (Fig. \ref{f_arcad_scorecard_3}). Conversely, GraphCast forecasts are closer to observations at lower levels, except for geopotential height. The dewpoint depression degradations suggest that further gains could be achieved by spectrally nudging the humidity field. However, this will only be viable once a fine-tuned version of GraphCast becomes available because of systematic differences between the moisture state estimates of ERA5 and the initializing GDPS analyses. Fine-tuning will bring the mean moisture state of GraphCast inferences closer to GDPS analyses and is expected to reduce errors associated with any adverse response from the GEM physics suite. The lower-level results also suggest that extending spectral nudging to the boundary layer could be beneficial. However, this would require an AI model trained on GEM's terrain-following vertical coordinate with more vertical levels in the boundary layer, fine-tuned to emulate the GDPS analyses for improved consistency with the GEM model’s lower-boundary forcings.

The small but growing GDPS-SN improvements over GraphCast in the mid-level layer for spectrally-nudged temperature and zonal wind are somewhat surprising (Fig. \ref{f_arcad_scorecard_3}). The nudging relaxation time of $\tau=12$ hr implies that the large scales in GDPS-SN are not too tightly constrained, which is visible in Fig. \ref{f_smooth}. This may allow for some large-scale improvements through upscale propagation of information from an improved fine-scale representation. However, this choice of $\tau$ also leads to a small averaging effect caused by the blending of two forecasts. To a lesser extent, smoothing attributable to this effect may also have contributed to the noted improvements.

\begin{figure}
\begin{center}
\noindent\includegraphics[width=28pc,angle=0]{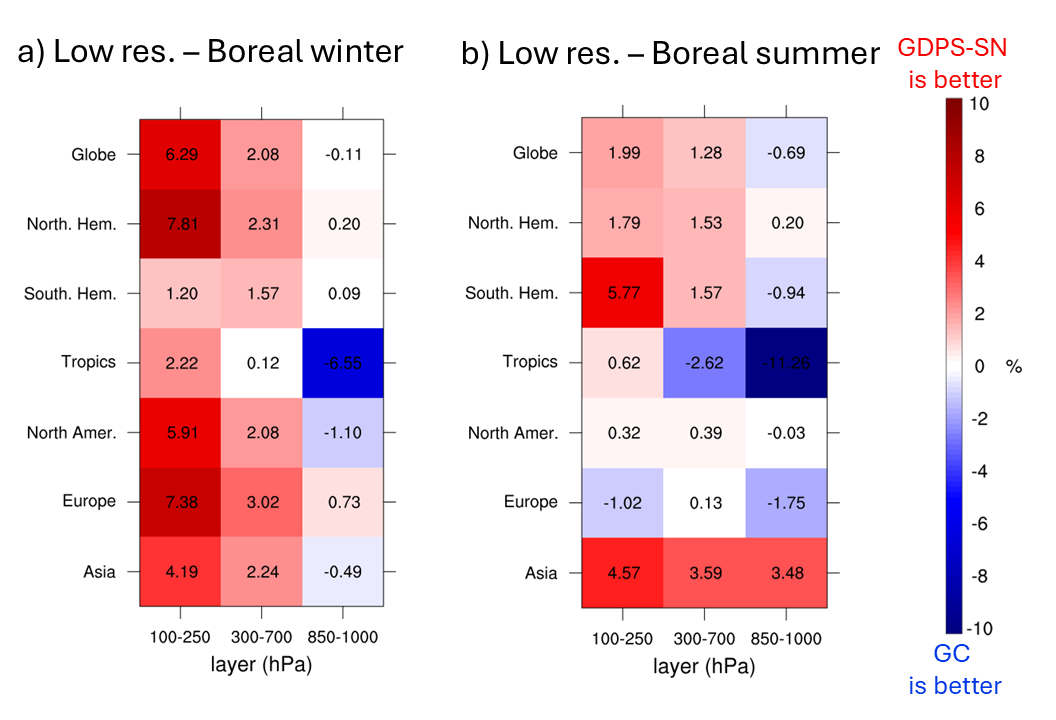}\\
\caption{Same as in Fig. \ref{f_heatmap_1}, except that red shadings indicate a reduction of RMSE by GDPS-SN with respect to GraphCast, whereas blue shadings indicate the opposite. Only verification using forecasts at low resolution is shown, but for both seasons.}
\label{f_heatmap_3}
\end{center}
\end{figure}

Finally, the regional comparison of GDPS-SN and GraphCast forecasts at low resolution (Fig. \ref{f_heatmap_3}) shows that the prediction accuracy of GDPS-SN is typically slightly better than or comparable to GraphCast, with the tropical boundary layer being one of the exceptions. Fig. \ref{f_heatmap_1} reveals that, compared to GraphCast, GDPS-CTL large scales have substantially reduced accuracy in the tropics, which is slightly improved with GDPS-SN (Fig. \ref{f_heatmap_2}). To leverage the full potential of GraphCast in the tropics it will likely be necessary to extend nudging in the boundary layer, once fine-tuning to GDPS analyses is complete.

\subsection{Verification against ECMWF analyses}
\begin{figure}
\begin{center}
\noindent\includegraphics[width=38pc,angle=0]{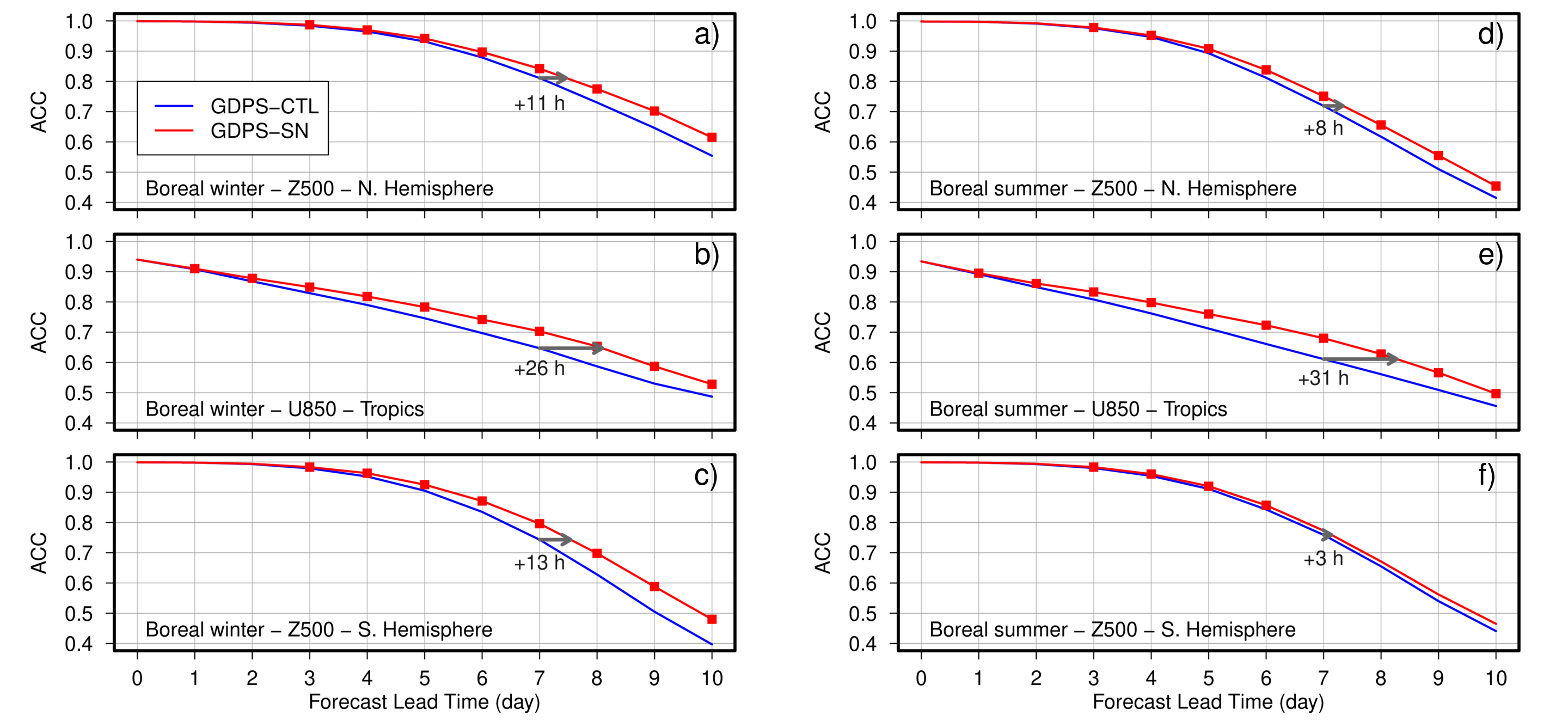}\\
\caption{Anomaly correlation coefficient (ACC) against the ECMWF analyses with GDPS-CTL (blue) and GDPS-SN (red) for (a,d) northern hemisphere 500-hPa geopotential height, (b,e) tropical 850-hPa zonal wind, and (c,f) southern hemisphere 500-hPa geopotential height. Results are presented for 60 cases of boreal winter 2022 (left column: a--c) and 62 cases of boreal summer 2022 (right column: d--f). ACC is computed using climatology of ERA-Interim reanalyses. Red and blue markers denote statistically significant differences in favor of GDPS-SN and GDPS-CTL, respectively. No marker for a forecast lead time implies that the null hypothesis, stating that the statistics of the two samples are the same, cannot be rejected based on the 95th percentile. Grey arrows with printed numbers depict gains in predictability in terms of forecast hours.}
\label{f_verdict_ano_cor_1}
\end{center}
\end{figure}

The comparison of GDPS-SN and GDPS-CTL forecasts against ECMWF analyses -- in terms of $f_{RMSE}$ -- leads to conclusions that are similar to those of the radiosonde assessment: general improvements for all variables, lead times, and pressure levels (not shown). Figure \ref{f_verdict_ano_cor_1} provides an example of changes in anomaly correlation coefficient (ACC). For 500-hPa geopotential heights, the ACC improvements with GDPS-SN (relative to GDPS-CTL) in day-7 forecasts over the Northern Hemisphere region are equivalent to a 11-hr increase in predictability during boreal winter and an 8-hr increase in summer (Figs. \ref{f_verdict_ano_cor_1}a,d). The corresponding ACC improvements over the Southern Hemisphere region are about 3 hours during boreal summer and 13 hours during boreal winter (Figs. \ref{f_verdict_ano_cor_1}c,f). There are also notable improvements in zonal wind ACC in the Tropics that are equivalent to 26 hours or more (Figs. \ref{f_verdict_ano_cor_1}b,e). The strength of GraphCast in improving the ACC around day 5 (and beyond) is thus well leveraged by the hybrid system. GraphCast has not been included in this figure because its ACC is strongly influenced by the excessive fine-scale smoothing. To fairly compare GraphCast against the two versions of GDPS, it is necessary to do so at low resolution (section \ref{s_rp}b); however, based on spectral coherence presented in Figs. \ref{f_var_coh_2}c--d, low-resolution ACC of GraphCast is expected to be comparable to GDPS-SN.

\subsection{Evaluation against surface observations}
For near-surface evaluation, combined SYNOP, METAR, and Surface Weather and marine OBservations (SWOB; available only over Canada and distributed by ECCC) data were used for surface pressure, screen-level temperature and dewpoint temperature, and anemometer-level wind speed. For precipitation, ground observations of 24-hr accumulation, subject to collection and quality control by the CAnadian Precipitation Analysis system (CaPA; \citealt{lfr15}), are used. Any data originating from stations with an altitude difference larger than 100 m with respect to GDPS orography is excluded, whereas any observed or forecast wind speed below 1.5 $\text{m s}^{-1}$ is set to 0.

\begin{figure}
\begin{center}
\noindent\includegraphics[width=28pc,angle=0]{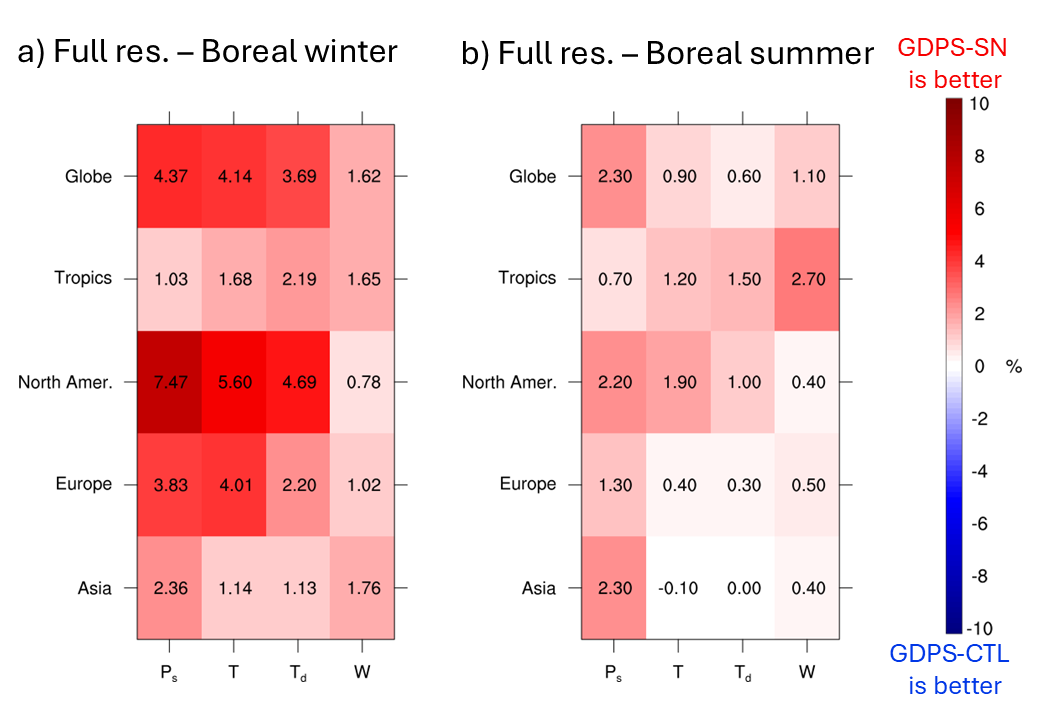}\\
\caption{Heatmap of the changes in the forecast quality index (Eq. \ref{e_rmse}) against combined SYNOP, SWOB, and METAR observations for surface pressure ($P_s$), screen-level temperature ($T$), dewpoint temperature ($T_d$), and anemometer-level wind speed ($W$) in various geographical domains for (a) 60 cases of boreal winter 2022 and (b) 62 cases of boreal summer 2022. Results are computed at full resolution. Red shadings indicate a reduction of the RMSE in GDPS-SN with respect to GDPS-CTL, while blue shadings indicate the opposite.}
\label{f_emet_score_1}
\end{center}
\end{figure}

\begin{figure}
\begin{center}
\noindent\includegraphics[width=38pc,angle=0]{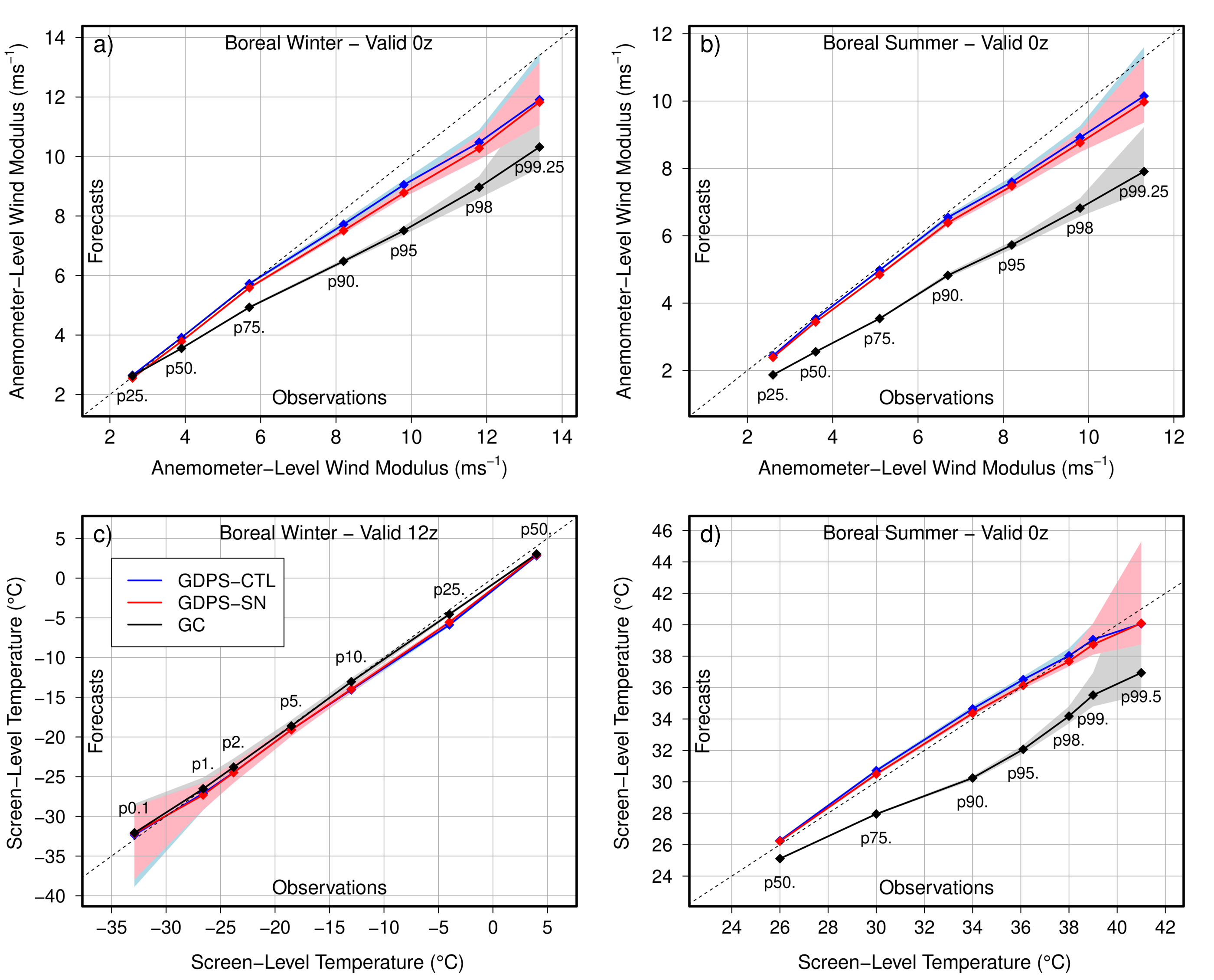}\\
\caption{Q-Q plots of day-5 GraphCast (black), GDPS-CTL (blue), and GDPS-SN (red) forecasts against combined SYNOP, SWOB, and METAR for: (a, b) anemometer-level wind speeds, and (c, d) screen-level temperatures over North America for (a, c) 30 cases of boreal winter valid at (a) 00 and (c) 12 UTC, and (b, d) 31 cases of boreal summer valid 00 UTC. Local standard time over North America is mostly between UTC-5 and UTC-8 h. Results correspond to full resolution of the different models. Diamond symbols denote selected percentiles with their values printed. Light shadings indicate the 5\%--95\% confidence interval for the inverse cumulative distribution functions, based on the Kolmogorov-Smirnov statistic.}
\label{f_qq_1}
\end{center}
\end{figure}

Guidance from GDPS-SN  is generally improved over GDPS-CTL for all variables and regions (Fig. \ref{f_emet_score_1}). The improvements are, however, smaller than in the upper air (Fig. \ref{f_heatmap_2}). Achieving considerable improvements near the surface appears to be difficult without introducing spectral nudging in the boundary layer. However, spectral nudging in the mid-level layer does have a considerable positive impact on the surface pressure via redistribution of upper-air mass. The improvements during boreal summer are reduced, consistent with the upper-air verification discussed above.

The ability of the GDPS-SN to represent high-impact weather (high wind speeds and extreme temperatures) is not significantly different from the GDPS-CTL (Fig. \ref{f_qq_1}). Both are substantially better than GraphCast, which is unable to predict the tails of the climatological distributions. For example, the distribution of late-afternoon wind speeds over North America provided by GraphCast shows a considerable shift toward lower values, which increases at high percentiles of the distribution, resulting in a -3 $\text{m s}^{-1}$ bias near the 99$^\text{th}$ percentile (Figs. \ref{f_qq_1} a, b). Although GraphCast does not show any discernible bias for low temperature percentiles over North America in winter (Fig. \ref{f_qq_1} c), it suffers from a 2$\sim$4 $\degree$C cold bias at mid-to-high temperature percentiles in summer (Fig. \ref{f_qq_1} d), a problem that does not affect GDPS-SN. At high temperature percentiles,  GraphCast’s poor performance with respect to the extremes can likely be explained by its lack of well-resolved fine scales \citep{ivs20} combined with inconsistencies in the surface forcing between ERA5 and GDPS analyses. As the spectral nudging configuration presented in this study only targets synoptic scales, the GEM component of the hybrid system is able to fill in small scales associated with local forcings. GDPS-SN is, therefore, not adversely affected by GraphCast’s limitations regarding the extremes.

\begin{figure}
\begin{center}
\noindent\includegraphics[width=38pc,angle=0]{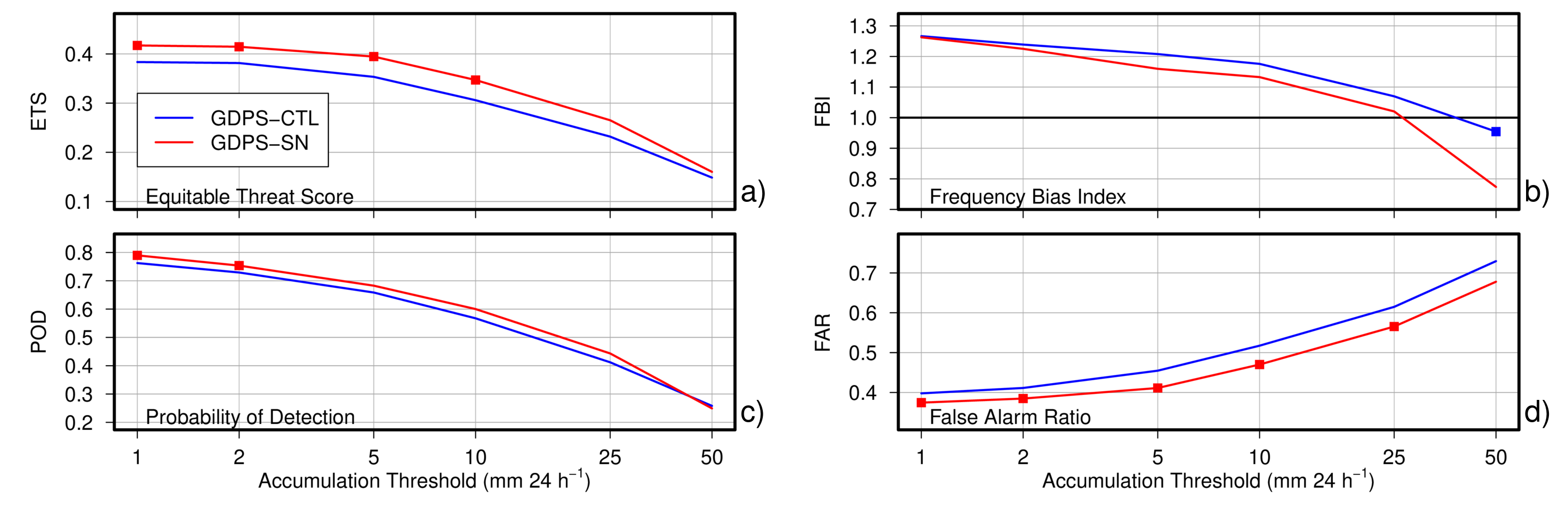}\\
\caption{Quantitative precipitation forecast verification for day 5 assessed by 24-hr precipitation threshold (accumulations between 108 and 132 hr of integration) with (a) equitable threat score (ETS), (b) frequency bias index (FBI), (c) probability of detection (POD), and (d) false alarm ratio (FAR). Results from the 60 GDPS-CTL (blue) and GDPS-SN (red) cases of boreal winter are compared to ground observations over North America used by CaPA. Red and blue line markers denote statistically significant differences in favor of GDPS-SN and GDPS-CTL, respectively. Significance is computed by bootstrapping 3-day data blocks consisting of forecast-observation pairs from all stations. No marker at a threshold level implies that the null hypothesis, stating that the statistics of the two samples are the same, cannot be rejected based on the 90th percentile.}
\label{f_precip_wint}
\end{center}
\end{figure}

Spectral nudging also significantly improves boreal winter precipitation guidance over North America primarily through a reduction in false alarm (Fig. \ref{f_precip_wint}). These improvements are likely the direct result of an improved representation of the synoptic-scale flow in strongly forced winter conditions. The impact of spectral nudging on summer precipitation over North America is, however, negligible (not shown). During summer, when weak synoptic forcings are prevalent, nudging only large scales is not expected to considerably improve precipitation forecasts, although extending nudging to the boundary layer may lead to some improvements by mitigating model biases.

\subsection{Tropical cyclone evaluation}
Evaluation of the impact of spectral nudging on the tropical cyclone (TC) guidance is performed using the data from the International Best Track Archive for Climate Stewardship (IBTrACS; \citealt{kkl10}) following the methodology described by \cite{mca24}.

\begin{figure}
\begin{center}
\noindent\includegraphics[width=37pc,angle=0]{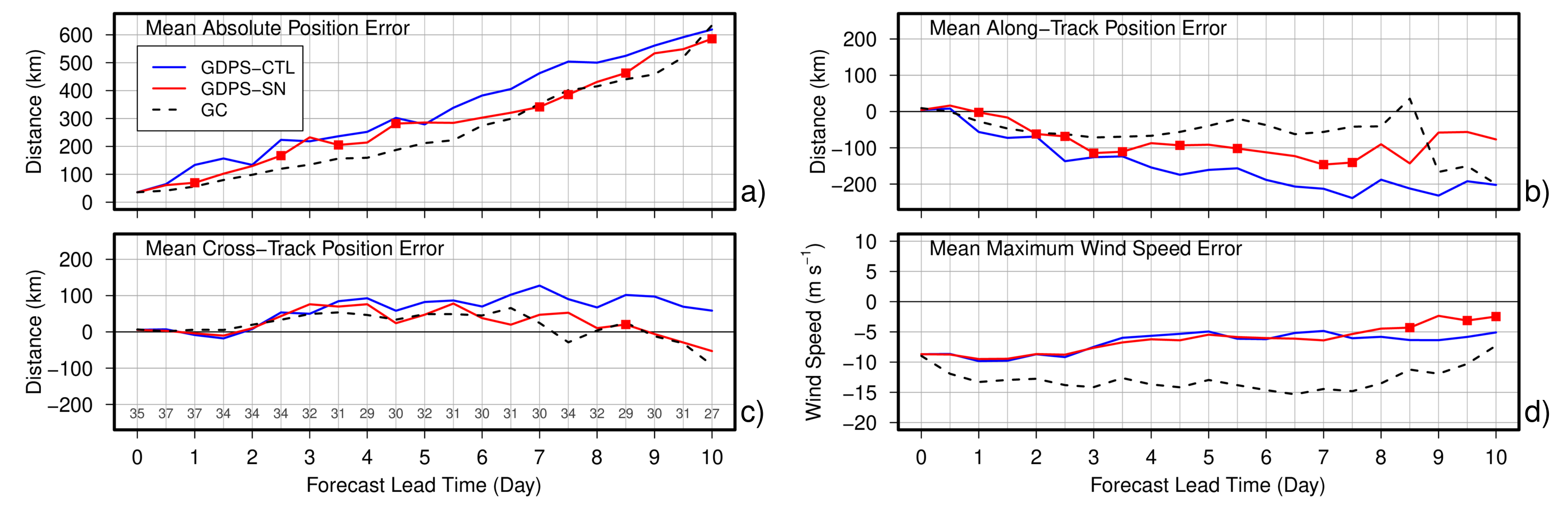}\\
\caption{Tropical cyclone tracking performance comparison of the GDPS-CTL (blue) and GDPS-SN (red) against IBTrACS data in three northern hemisphere basins (West Atlantic, East, and West Pacific) for 62 cases of boreal summer 2022: (a) mean absolute position error, (b) mean along-track position error, (c) mean cross-track position error, and (d) mean maximum wind speed error. Performance of GraphCast (dashed black) is presented for reference. Negative along-track errors imply cyclones move too slowly and positive cross-track errors imply cyclones veer too much to the right with respect to the observed trajectory. Red and blue markers denote statistically significant differences in favor of GDPS-SN and GDPS-CTL, respectively. No marker at a lead time implies that the null hypothesis, stating that the statistics of the two samples are the same, cannot be rejected based on the 90th percentile. The number of cases for different forecast lead times are indicated in \ref{f_tc}c.}
\label{f_tc}
\end{center}
\end{figure}

AI-based models, including GraphCast, are generally capable of predicting the TC trajectories with enhanced accuracy \citep{lsw22}. Results presented in Fig. \ref{f_tc}a show that spectral nudging allows GDPS-SN to leverage GraphCast's enhanced TC steering accuracy and leads to an overall reduction in position error in the predicted TC trajectories. Notably, GDPS-CTL's tendency of predicting storms that move too slowly -- indicated by negative along-track error -- is improved with GDPS-SN, especially at lead times beyond day 5 (Fig. \ref{f_tc}b). There is also an indication of improvement with respect to GDPS-CTL's tendency of predicting TCs that veer too much to the right from their observed tracks, which is indicated by positive cross-track error (Fig. \ref{f_tc}c). Conversely, there is little or no significant impact of spectral nudging on TC intensity, as measured by the maximum sustained wind speed (Figs. \ref{f_tc}d). Although both GDPS-CTL and GDPS-SN have a similar weak-intensity bias, which is a well-known weakness of the GEM model (see \cite{mca24}, for more details), both versions of GDPS nevertheless predict higher intensity TCs compared to GraphCast. Severe TC weak-intensity biases are typical of state-of-the-art deterministic AI-based forecast models due to the double-penalty effect caused by learning to minimize the MSE \citep{bac24}. Combined with the small scales of the TC vortex that make direct nudging of the pertinent scales infeasible, this means that the potential for improvements in intensity prediction lies primarily in the physics-based NWP component of the hybrid system.

\subsection{Computational cost of spectral nudging}
\label{s_eva_cc}
The current implementation of spectral nudging in GEM with the presented optimal configuration leads to an increase of computational cost by approximately 25\%. This is in addition to the relatively insignificant cost of generating the GraphCast inferences. The nudging overhead scales linearly with the number of fields and levels to which the technique is applied. Due to the proof-of-concept nature of this study, no optimization of the informatics code has been performed. In the future, this cost could be considerably reduced through code optimization for the spectral filter and implementation of an asynchronous input server. Further improvements could be attained by implementing time-varying nudging, where $\tau$ is allowed to change over time following a cosine-bell profile as proposed by \cite{hsy14}. With this approach, the application of spectral nudging could be restricted to model time steps close to the inference times of GraphCast, potentially leading to significant computational cost savings. However, these considerations lie beyond the scope of this study.

\section{Summary and Future Work}
\label{s_summ}
The emergence of AI-based weather prediction models has disrupted the operational paradigm long dominated by the physics-based systems. The pressing question that motivates this study is whether it is possible to improve guidance from operational NWP models by leveraging the predictive skill of AI inferences.

A careful comparison of physics-based GEM and AI-based GraphCast predictions reveals that the latter suffers from excessive smoothing up to synoptic scales. At larger scales, however, GraphCast predictions are found to be highly skillful. Inspired by the improved large-scale skill of GraphCast, a hybrid NWP-AI system -- namely, GDPS-SN -- has been developed to produce global forecasts in which GEM's large-scale state is spectrally nudged toward larges scales of GraphCast inferences. This hybrid system is capable of generating real-time forecasts with accuracy that significantly surpasses ECCC's operational GPDS. The RMSE of the 500-hPa geopotential height is reduced by 5-10\%, with the largest predictability improvements attained around day 7 that exceed 24 hours over the tropics. This accurate prediction of large-scale circulation improves tropical cyclone steering estimates and the associated track predictions.

Although these results are achievable directly with AI-based systems, the hybrid model also generates the full spectra of fine scales that represents the tails of the climatology. This allows GDPS-SN to predict weather extremes that are challenging for the current generation of purely ``data-driven'' models.

Operational NWP models generate hundreds of internally and physically consistent forecast fields at high vertical resolution and temporal frequency. These outputs serve as essential guidance for operational meteorologists, especially for forecasts of high-impact events. The proposed hybrid system maintains these necessary capabilities, with a reasonable increase in computational cost. Increasing global AI model resolution and expanding the number of predicted variables represent real computational challenges for both training and inference. Moreover, the procedures needed to create data-driven predictions of the multitude of unanalyzed variables in a physically consistent way has not yet been devised. In this regard, the proposed hybrid system compensates for the perceived weaknesses of NWP models while addressing the limitations of current AI models.

The first version of GDPS-SN became operational (with an experimental status) at ECCC in March 2025. It is based on a GraphCast version that has been retrained and fine-tuned at ECCC. Work is also in progress to fine-tune GraphCast to emulate operational GDPS analyses. This fine-tuned version is expected to enhance GraphCast's skill with the 37 pressure-level version, particularly in the stratosphere and the boundary layer. Any such improvement will quickly be integrated into the hybrid GDPS-SN system.

Although this study used the GEM and GraphCast models for hybridization, the protocol established here could be applied to any pair of systems, provided that the physical model supports spectral nudging. Hybrid systems like the one described here may represent an optimal blending of the individual advantages of the physics- and data-based approaches to weather prediction in the foreseeable future.

The results from this study imply that, rather than viewing AI and NWP models as two competing paradigms, it is likely more prudent to consider these forecasting methods as complementary. A well-designed fusion of these two approaches can significantly mitigate their individual limitations while allowing for harnessing their respective strengths to provide better meteorological guidance. Therefore, instead of focusing on efforts to replace one with the other, future research should prioritize improving both NWP and AI models.

\acknowledgments
The authors thank their retired colleague, Dr. Bertrand Denis, for speculating about the potential benefits of spectrally nudging NWP models toward AI inferences, and Benoit Archambault for developing a vital tool that converts GraphCast outputs to ECCC's standard file format.


\appendix[A]
\appendixtitle{Spectral decomposition of forecast activity}
Let $x$ and $y$ denote forecast and analysis, respectively. The forecast and analysis climatological means are defined as $x_c =E [x]$ and $y_c = E[y]$, respectively, where $E$ denotes the expectation. In general, the climatology is a function of the date and time of the year. The corresponding anomalies are defined as $x'=x-x_c$ and $y'=y-y_c$.

The \emph{activity} of forecast and analysis can then be defined, respectively, as
\begin{eqnarray}
\label{e_Def_A}
A_x &= \sqrt{S_{x',x'}},\\
A_y &= \sqrt{S_{y',y'}} ,
\end{eqnarray}
where, given any two 2-D scalar fields $u$ and $v$, the operator $S$ is defined as
\begin{equation}
\label{e_Def_S}
S_{u,v} \equiv E \bigg [\Big \langle{(u-\langle{u}\rangle)(v-\langle{v}\rangle ) } \Big \rangle \bigg ].
\end{equation}
Here, the angle brackets denote spatial averaging, including latitudinal weighting.

The activity ratio of the forecast, normalized by the analysis activity, can then be obtained as follows:
\begin{equation}
\label{e_Gamma_total}
\Gamma =\sqrt{\frac{S_{x',x'}}{S_{y',y'}}},
\end{equation}
and the anomaly correlation coefficient is given as
\begin{equation}
\label{e_Def_ACC}
P =\frac{S_{x',y'}}{\sqrt{S_{x',x'}S_{y',y'}}}.
\end{equation}
These two coefficients quantify, respectively, the amplitude and phase errors of forecast anomalies with respect to the climatological mean.

By expressing the forecast and analysis fields as truncated spherical harmonic expansions, the covariance can be decomposed into contributions from individual harmonic components. Applying Parseval's identity and grouping terms by zonal wavenumber, $S$ can be written as a sum over total spherical wavenumbers $n$ as follows:
\begin{equation}
\label{e_Def_spec}
S_{u,v} = \sum_{n=1}^{N_{trunc}} E \big [ \sigma_{u,v}(n) \big ],
\end{equation}
where $\sigma_{u,v}(n)$ denotes the cross spectral density between $u$ and $v$ at wavenumber $n$. When $u=v$, $\sigma_{u,u}$ corresponds to the power spectral density of $u$. This allows for introducing a spectral activity ratio as a function of wavenumber $n$ as
\begin{equation}
\label{e_Gamma_PSD}
\gamma(n) =\sqrt{\frac{E \big [ \sigma_{x',x'}(n) \big ]}{E \big [ \sigma_{y',y'}(n)\big ]}} ,
\end{equation}
and a spectral anomaly correlation coefficient, hereafter referred to as spectral coherence, as
\begin{equation}
\label{e_Rho_PSD}
\rho(n) =\frac{E \big [ \sigma_{x',y'}(n) \big ]}{\sqrt{E \big [ \sigma_{x',x'}(n)\big ] \,E \big [ \sigma_{y',y'}(n)\big ]}} .
\end{equation}

\begin{figure}
\begin{center}
\noindent\includegraphics[width=39pc,angle=0]{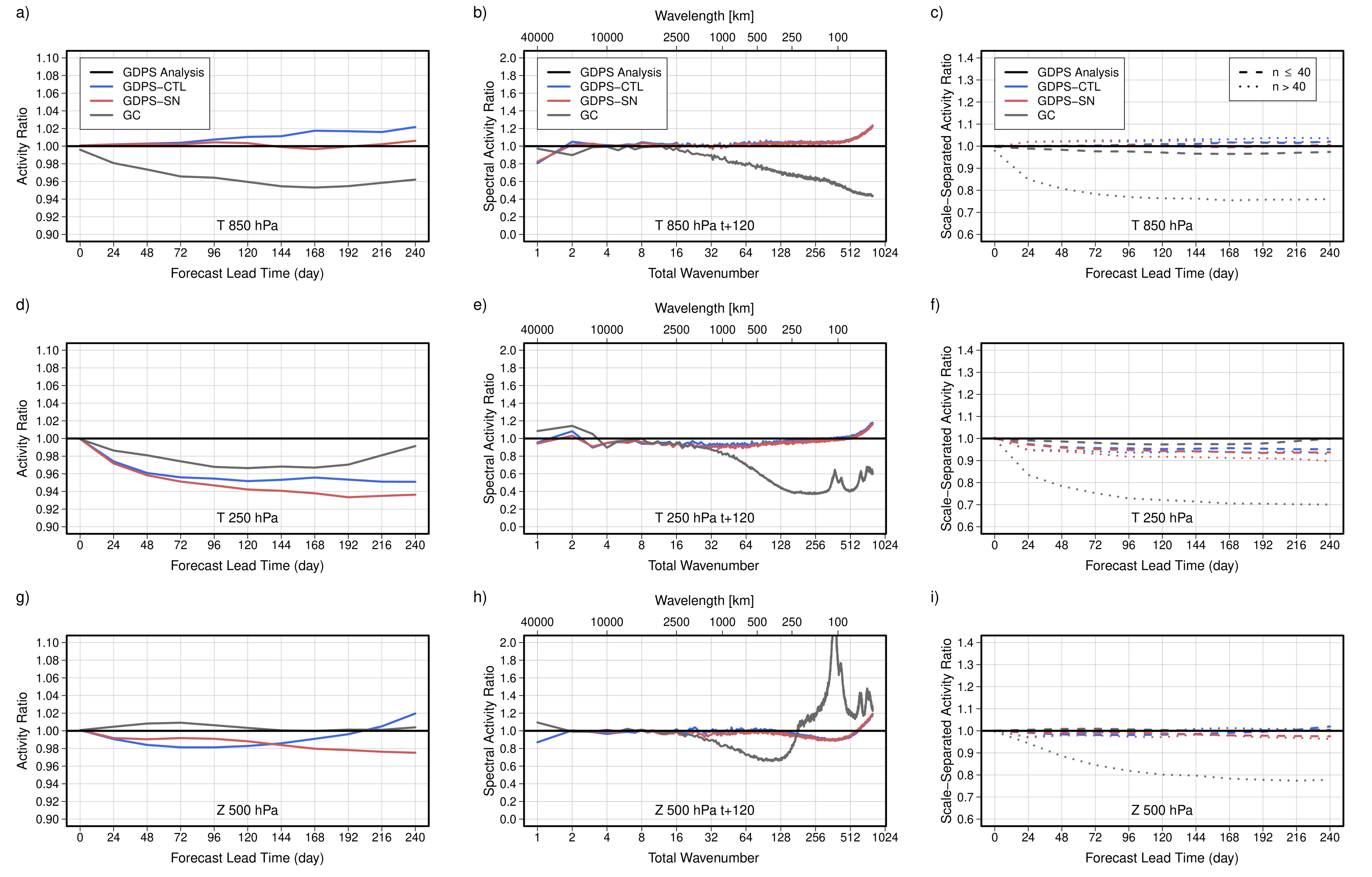}\\
\caption{Normalized activity (left column), spectral activity ratio (middle column) at day 5, and scale-separated normalized activity ratio (right column) for temperature at 850 hPa (top row) and 250 hPa (middle row), and geopotential at 500 hPa (bottom row) for GDPS-CTL (blue), GDPS-SN (red), and GraphCast (grey). Forecast activity is calculated using ECMWF analysis climatology, whereas normalization is done using GDPS analyses. Results correspond to boreal winter.}
\label{f_c1}
\end{center}
\end{figure}

Estimating model climatology $x_c$ requires running hindcasts, which may be impractical outside of an operational NWP context. The conventional workaround is to use a common climatology database, setting $x_c=y_c=c$. The results presented in Figs. \ref{f_c1}a--i correspond to this approach, where anomalies are calculated relative to a common ERA-Interim reanalysis climatology $c$ \citep{dus11}.

Alternatively, approximate climatological means $x_c$ and $y_c$ can be obtained as $x_c=\overline{x}$ and $y_c=\overline{y}$, where the overbar denotes average over a set of cases for a given season. This is the approach adopted throughout this paper. Therefore, Eqs. 1 and 2 are equivalent to Eqs. \ref{e_Gamma_PSD} and \ref{e_Rho_PSD}, respectively, where expectations are estimated by averaging spectral densities over a set of cases for a given season. Both approaches deliver qualitatively similar insights regarding model performance at different scales (not shown).

Figure \ref{f_c1}a illustrates the normalized activity ratio for temperature at 850 hPa across different lead times for predictions from GDPS-CTL, GDPS-SN, and GraphCast. By day 5, GraphCast's activity is suppressed by about 4\% compared to GDPS analyses. GDPS-CTL and GDPS-SN show similar activity, with GDPS-CTL being slightly overactive (up to 2\%) beyond day 5. The spectral activity ratio at day 5 (Fig. \ref{f_c1}b) shows increasingly suppressed activity for GraphCast at large wavenumbers, starting from wavenumber 16 (approx. 2500 km). For scales below 100 km, activity reduction is nearly 50\%. However, this significant reduction in fine-scale activity is not reflected in the total activity ratio (Fig. \ref{f_c1}a). Separating normalized activity for scales larger and smaller than 1000 km (wavenumber $<$40 and $>$40) in Fig. \ref{f_c1}c shows that total activity ratio is mainly reflective of scales larger than 1000 km. This inference is based on the fact that the total activity ratios (dashed lines in Fig. \ref{f_c1}c) almost match large-scale activity ratios (solid lines in Fig. \ref{f_c1}a), while fine-scale activity ratios (dotted lines in Fig. \ref{f_c1}c) indicate significant suppression in GraphCast.

Figures \ref{f_c1}d--f further demonstrate the limitation of total activity ratio in identifying fine-scale smoothing effects. For temperature at 250 hPa, GraphCast's total activity is the highest among models (Fig. \ref{f_c1}d) despite a 60\% reduction in activity for scales below 500 km (Fig. \ref{f_c1}e). It turns out that the total activity metric rewards GraphCast for over-predicting activity at the planetary scales (wavenumber$<$4) and is insensitive to an overall 30\% deficiency in activity for scales smaller than 1000 km (Fig. \ref{f_c1}f).

Figures \ref{f_c1}g–-i illustrate, for 500 hPa geopotential height, that activity as a metric can also be insensitive to spurious fine-scale variance. Specifically, both total activity (Fig. \ref{f_c1}g) and fine-scale activity (Fig. \ref{f_c1}i) for GraphCast fail to capture the unphysical increase in fine-scale variance concentrated near wavenumber 400 (Fig. \ref{f_c1}h)—a known issue with GraphCast, as previously reported by \cite{lsw23}. Overall, these results highlight the limitations of total activity for assessing mesoscale issues in model predictions (smoothing or spurious variance) and underscores the importance of its spectral decomposition for proper assessment of model performance.

Finally, for all cases shown in Fig. \ref{f_c1}, GDPS-SN's fine-scale activity is close to GDPS-CTL, exhibiting only slight suppression depending on the specific variable and level. This behavior is consistent with the spectral activity and amplitude ratios shown in Figs. \ref{f_c1} and \ref{f_var_coh_2}, respectively.

\begin{figure}
\begin{center}
\noindent\includegraphics[width=32pc,angle=0]{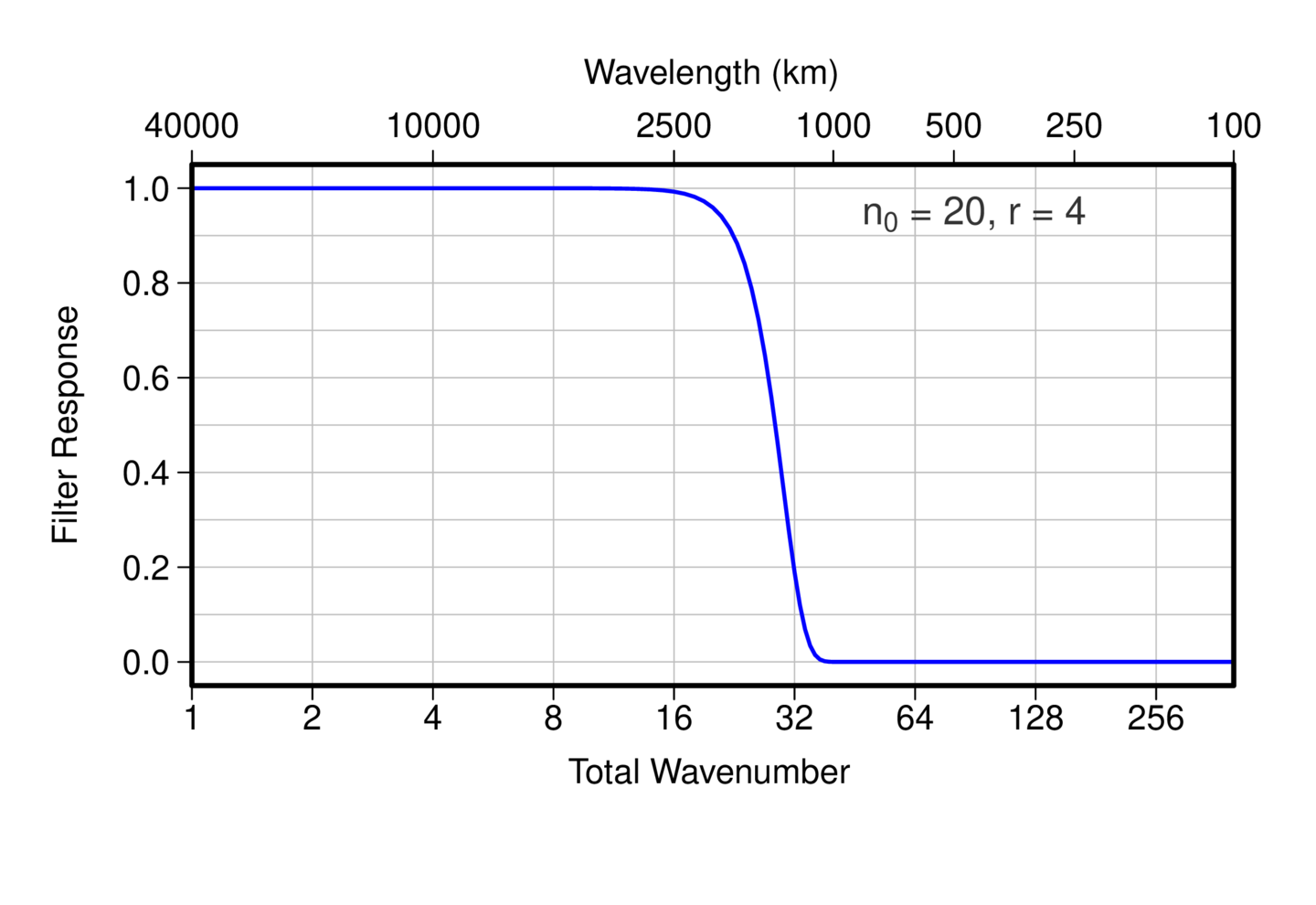}\\
\caption{Response of the spherical-harmonics based spectral filter with $n_0=30$ and $r=4$ (see Eq. \ref{e_a_filt}).}
\label{f_a1}
\end{center}
\end{figure}

\appendix[B]
\appendixtitle{The spherical harmonics-based global filter}
Isolation of the large scales in the predictions from GDPS and GraphCast is required for a fair comparison of forecast accuracy. This is achieved through the application of a spectral filter. The physical outputs from the models at a given pressure level are first transformed to the spectral space through spherical harmonic-based decomposition. This is followed by the application of the filter, $f_n$, as proposed by \cite{sho84}. The functional form of the filter is given by

\begin{equation}
\label{e_a_filt}
  f_n = \exp\left[-\left(\frac{n(n+1)}{n_o(n_o+1)}\right)^{r\;} \right],
\end{equation}

\noindent where $n$ denotes the total wavenumber, $n_o$ is the cut-off total wavenumber, and the exponent $r$ is a non-dimensional parameter related to the sharpness of the filter response. Based on the spectral comparison of the GDPS and GraphCast predictions, the filter is configured by setting $n_0=30$ and $r=4$. As shown in Fig. \ref{f_a1}, the resulting filtered fields fully retain amplitudes for scales associated with approximately $n<20$ (wavelength$>$2000 km) and fully removes scales corresponding to $n>40$ (wavelength$<$1000 km). The physical fields reconstructed from the filtered spectra leads to the desired filtered fields.

\appendix[C]
\appendixtitle{The DCT-based filter for spectral nudging}
Two-dimensional DCT and its implementation in GEM for spectral nudging in limited-area modelling are well documented in the existing literature \citep{dcl02,hsy14}. For global simulations based on the Yin-Yang grid system, nudging is applied to the LAM sub-domains associated with the Yin and Yang grids separately. The first step in spectral nudging is to compute the coefficients, $\hat{f}(m,n)$, of the DCT of $(F_{GC}-F_{GEM})$, which determines the nudging increment (see Eq. \ref{e_nudge}). For each of the Yin-Yang sub-domains, $m$ and $n$ denote the one-dimensional horizontal wavenumbers. In the second step,  $\hat{f}(m,n)$ is subjected to a spectral filter, ${f_F}(m,n)$, of the form

\begin{equation}
\label{e_b_filt}
     f_F(m,n)=
\begin{cases}
    0.0,             & \text{if } \hat{\alpha}>\lambda_{LS}/\lambda_{SS}\\
    \left[\cos\left(\frac{\pi}{2} \frac{\alpha\lambda_{LS}/(2\Delta) -1}{\lambda_{LS}/\lambda_{SS}-1}\right) \right]^2,& \text{if } 1.0<\hat{\alpha}\leq \lambda_{LS}/\lambda_{SS}\\
    1.0,             & \text{if } \hat{\alpha}\leq1.0
\end{cases}
\end{equation}

\noindent where $\Delta$ denotes the model grid spacing, $\alpha$ is the normalized two-dimensional wavenumber given by $\alpha=\sqrt{\frac{m^2}{N_i^2}+\frac{n^2}{N_j^2}}$ associated with each of the Yin-Yang sub-domains of size $(N_i\times N_j)$, and $\hat{\alpha}=\alpha\lambda_{LS}/(2\Delta)$.

\begin{figure}
\begin{center}
\noindent\includegraphics[width=39pc,angle=0]{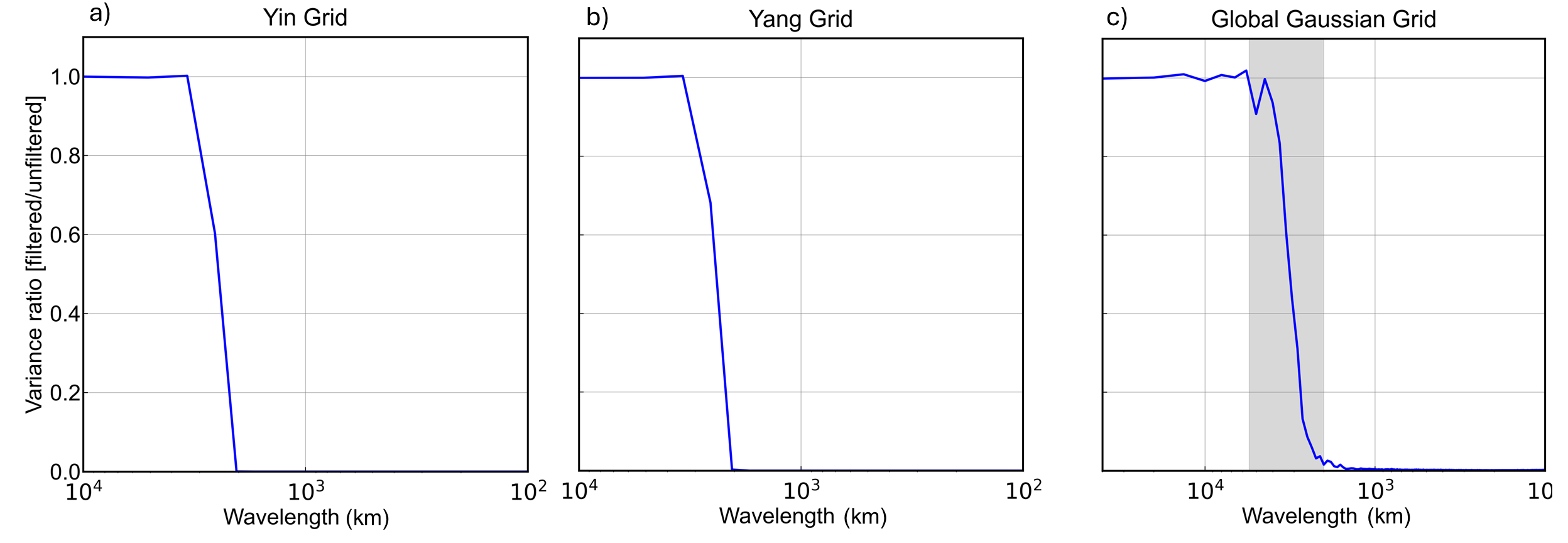}\\
\caption{Response of the DCT-based filter based on $\lambda_{LS}=2750$ km and $\lambda_{SS}=2250$ km (see Eq. \ref{e_b_filt}) with respect to the (a) Yin grid, (b) Yang grid, and (c) a global Gaussian grid for temperature at 500 hPa. Over the global Gaussian grid, the small- and large-scale cut-offs for the optimal configuration are approximately 2000 km and 5500 km, respectively, as indicated with the shaded area.}
\label{f_b1}
\end{center}
\end{figure}

Figures \ref{f_b1}a and \ref{f_b1}b illustrate the response of this DCT-based filter for the optimal configuration based on $\lambda_{LS}=2750$ km and $\lambda_{SS}=2250$ km. The figures demonstrate that the DCT-based filter is capable of targeting the desired scales for the individual Yin-Yang sub-domains. The global response of the filter was determined by first interpolating a field (e.g., temperature) from the Yin-Yang grid to a global Gaussian grid before and after applying the DCT-based filter. The spectral variance ratio between the filtered and unfiltered fields was then computed using a spherical harmonics-based decomposition. The corresponding results for temperature at 500 hPa are presented in Fig. \ref{f_b1}c for illustration.

\bibliographystyle{ametsocV6}
\bibliography{references_nwp_ai_sn_s3}

\end{document}